\documentclass[fleqn,usenatbib]{mnras}

\usepackage{mathptmx}
\usepackage{float}

\usepackage[T1]{fontenc}
\usepackage{ae,aecompl}

\usepackage{graphicx}	% Including figure files
\usepackage{amsmath}	% Advanced maths commands
\usepackage{amssymb}	% Extra maths symbols
\usepackage[normalem]{ulem}
\usepackage{soul}
\usepackage{lmodern}

\newcommand{\LA}{Lyman-$\alpha$ }	% per cm-squared

\newcommand{\field}{{\hat \phi}}
\newcommand{\de}[1]{\partial_{#1 }}
\newcommand{\sumnn}{\sum_{j\in \text{NN}(i)}}

\newcommand{\G}{{\small P-GADGET}3~}
\newcommand{\AG}{{\small AX-GADGET}~}

\renewcommand{\vec}[1]{\mathbf{#1}}

\title[Lyman-$\alpha$ and LSS properties in FDM cosmologies]{Lyman-$\alpha$ forest and non-linear structure characterization in Fuzzy Dark Matter cosmologies}

\author[M. Nori et al.]{
Matteo Nori,$^{1,2,3}$\thanks{E-mail: matteo.nori3@unibo.it}
Riccardo Murgia,$^{4,6}$
Vid Ir\v{s}i\v{c},$^{5}$
Marco Baldi$^{1,2,3}$
and Matteo Viel$^{4,6,7}$
\\
$^{1}$Dipartimento di Fisica e Astronomia, Alma Mater Studiorum - University of Bologna, Via Piero Gobetti 93/2, 40129 Bologna BO, Italy\\
$^{2}$INAF - Osservatorio Astronomico di Bologna, Via Piero Gobetti 93/3, 40129 Bologna BO, Italy\\
$^{3}$INFN - Istituto Nazionale di Fisica Nucleare, Sezione di Bologna, Viale Berti Pichat 6/2, 40127 Bologna BO, Italy\\
$^{4}$SISSA, Via Bonomea 265, 34136 Trieste, Italy\\
$^{5}$University of Washington, Department of Astronomy,
3910 15th Ave NE, WA 98195-1580 Seattle, USA \\
$^{6}$INFN - Istituto Nazionale di Fisica Nucleare, Sezione di Trieste, Via Bonomea 265, 34136 Trieste, Italy\\
$^{7}$INAF - Osservatorio Astronomico di Trieste, via Tiepolo 11, I-34143 Trieste, Italy
}

\date{Accepted XXX. Received YYY; in original form ZZZ}

\pubyear{2018}
\begin{document}
\label{firstpage}
\pagerange{\pageref{firstpage}--\pageref{lastpage}}
\maketitle

\begin{abstract}
Fuzzy Dark Matter (FDM) represents an alternative and intriguing description of the standard Cold Dark Matter (CDM) fluid, able to explain the lack of direct detection of dark matter particles in the GeV sector and to alleviate small scales tensions in the cosmic large-scale structure formation. Cosmological simulations of FDM models in the literature were performed either with very expensive high-resolution grid-based simulations of individual haloes or through N-body simulations encompassing larger cosmic volumes but resorting on significant approximations in the FDM non-linear dynamics to reduce their computational cost. With the use of the new N-body cosmological hydrodynamical code \AG, we are now able not only to overcome such numerical problems, but also to combine a fully consistent treatment of FDM dynamics with the presence of gas particles and baryonic physical processes, in order to quantify the FDM impact on specific astrophysical observables. In particular, in this paper we perform and analyse several hydrodynamical simulations in order to constrain the FDM mass by quantifying the impact of FDM on Lyman-$\alpha$ forest observations, as obtained for the first time in the literature in a N-body setup without approximating the FDM dynamics. We also study the statistical properties of haloes, exploiting the large available sample, to extract information on how FDM affects the abundance, the shape, and density profiles of dark matter haloes.
\end{abstract}


%%%%%%%%%%%%%%%%%%%%%%%%%%%%%%%%%%%%%%%%%%%%%%%%%%

%%%%%%%%%%%%%%%%% BODY OF PAPER %%%%%%%%%%%%%%%%%%

\section{Introduction}

In the first half of the last century, the scientific community consensus gathered around two crucial facts about our Universe, that are now considered the pillars of modern cosmology: firstly that the Universe is expanding and it is doing so at an accelerated rate and, secondly, that the estimated baryonic matter content within it cannot account for all the dynamical matter needed to explain its gravitational behaviour. 

The standard cosmological framework built upon these concepts, called $\Lambda$CDM, still holds today. It implies the existence of \textit{dark energy}, as a source of energy for the accelerated expansion of the Universe, and of \textit{dark matter}, as an additional gravitational source alongside standard matter, without however specifying their fundamental nature that still represents a major puzzle for cosmologists.

The evidence for a \textit{cold} and \textit{dark} form of matter (CDM) --~a not-strongly electromagnetically interacting particle or a gravitational \textit{quid} that mirrors its effect~-- span over different scales and are related to dynamical properties of systems, as e.g. the inner dynamics of galaxy clusters \citep[][]{Zwicky_1937,Clowe06} and the rotation curves of spiral galaxies \citep[][]{Rubin_Ford_Thonnard_1980,Bosma_1981,Persic96}, but also to the gravitational impact on the underlying geometry of space-time, as strong  gravitational lensing of individual massive objects \citep[][]{Koopmans02} as well as the weak gravitational lensing arising from the large-scale matter distribution \citep[][]{Mateo98,Heymans_etal_2013,Planck_2015_Gravitational_Lensing,Hildebrandt_etal_2017}. Further evidence is based on the relative abundance of matter with respect to the total cosmic energy budget required in order to reconcile Large Scale Structures (LSS) --~as observed through low-redshift surveys~-- with the angular power spectrum of CMB temperature anisotropies that seed the early universe density perturbations \citep[as observed e.g. from WMAP and Planck][respectively]{wmap7,Planck_2015_XIII}, on the clustering of luminous galaxies \citep[see e.g.][]{VIPERS_Om_M,SDSS-III-final}, on the abundance of massive clusters \citep[][]{Kashlinsky98} and their large-scale velocity field \citep[][]{Bahcall98}.

Whether dark matter consists indeed of a yet undetected fundamental particle or it represents an indirect effect of some modification of Einstein's General Relativity theory of gravity is still widely debated. 
Nevertheless, it has been possible to exclude some of the proposed dark matter effective models, such as e.g. the Modified Newtonian Dynamics and its variants \citep[MOND see e.g.][]{Milgrom_1983, Sanders02, Bekenstein04}, recently ruled out \citep[][]{Chesler17} by the implications of the gravitational wave event GW170817 \citep[][]{GW170817}. The lack of detection of dark matter particles in the GeV mass range through neither of indirect astronomical observations \citep[see e.g.][]{Fermi17annih},  direct laboratory detections \citep[see e.g.][]{Danninger17}, nor artificial production in high-energy collisions experiments \citep[see e.g.][]{Buonaura18} has been undermining the appeal of the most massive dark matter particle candidates, as e.g. the Weakly Interactive Massive Particles (WIMPs), and it is presently shifting the scientific community efforts in the hunt of direct observations from such high mass ranges towards lower ones \citep[see e.g.][]{Bertone_Hooper_Silk_2005}.

A good starting point where to focus research and to clarify such long-standing uncertainties would be the apparent failures of the $\Lambda$CDM model at scales $\lesssim 10\ kpc$ --~as given e.g. by the cusp-core problem \citep[][]{Oh11}, the missing satellite problem \citep[][]{Klypin99}, the too-big-to-fail problem \citep[][]{Boylan-Kolchin12}~--, all arising as an apparent inconsistency between simulations and observations, the latter being more in line with less pronounced density fluctuations at those scales than predicted by the former. However, the nature of such apparent failures has been subject of debate in the astrophysics community. It is still unclear, in fact, whether they should be ascribed to an imperfect baryonic physics implementation in numerical simulations \citep[see e.g.][]{Maccio12,Brooks13}, to an intrinsic diversity of properties related to the formation history and local environment of each individual dark matter halo \citep[][]{Oman_etal_2015}, to the fundamental nature of the dark matter particle \citep[see e.g.][]{Spergel00,Rocha13,Kaplinghat00,Medvedev14} or even to a combination of all these possible causes.

Among the particle candidates that have been proposed in the literature, Fuzzy Dark Matter (FDM) models describe dark matter as made up of very light bosonic particles \citep[see e.g.][for a review on the topic]{Hui16}, so light that their quantum nature becomes relevant also at cosmological scales. This requires a description of dark matter dynamics in terms of the Schr\"odinger equation, in order to take into account quantum corrections, and can be mapped in a fluid-like description where a \textit{quantum potential} (QP) enters the classical Navier--Stokes equation \citep{Hu00}.

The typical wave-like quantum behaviour adds to the standard CDM dynamics a repulsive effective interaction that, along with creating oscillating interference patterns, actively smooths matter over-densities below a redshift-dependent scale that decreases with the cosmic evolution --~as confirmed by FDM linear simulations \citep[see e.g.][]{Marsh10,axionCAMB}~-- thus potentially easing some of the previously mentioned small-scale inconsistencies of the CDM model. 

The lack of density perturbations at small scales induced by the QP is represented, in Fourier space, by a sharp suppression of the matter power spectrum, that persists --~at any given scale~-- until the action range of the QP shrinks below such scale and cannot balance any longer the effect of the gravitational potential \citep[see e.g.][for another detailed review on the subject]{Marsh16}. As a matter of fact, while linear theory predicts that perturbations at scales smaller than the cut-off scale never catch up with those at larger scales --~untouched by FDM peculiar dynamics~--, non-linear cosmological simulations have shown that gravity is indeed able to restore intermediate scales to the unsuppressed level, in a sort of \textit{healing} process \citep{Marsh16nl,Nori18}.

FDM non-linear cosmological simulations have been performed over the years either with highly numerically intensive high-resolution Adaptive Mesh Refinement (AMR) algorithms able to solve the Schr\"odinger-Poisson equations over a grid \citep[see e.g.][]{GAMER,GAMER2} or with standard N-Body codes that, however, include the (linear) suppression only in the initial conditions but neglect the integrated effect of the FDM interaction during the subsequent dynamical evolution \citep[see e.g.][]{Schive16,Irsic17,Armengaud17} --~basically treating FDM as standard dark matter with a suppressed primordial power spectrum, similarly to what is routinely done in Warm Dark Matter simulations \citep[][]{Bode00}. The former approach led to impressive results in terms of resolution \citep[see e.g.][]{Woo09,Schive14} but is extremely computationally demanding, thereby hindering the possibility of adding a full hydrodynamical description of gas and star formation for cosmologically representative simulation domains. On the other hand, the latter allows for such possibility because of its reduced computational cost which is, however, gained at the price of the substantial approximation of neglecting QP effects during the simulation \citep[see e.g.][]{Schive16}.

For these reasons, following the approach first proposed in \citet{Mocz15}, we devised \AG \citep{Nori18}, a modified version of the N-body hydrodynamical cosmological code \G \citep{Springel05}, to include the dynamical effect of QP through Smoothed Particle Hydrodynamics (SPH) numerical methods. The explicit approximation of the dependence on neighbouring particles results in a less numerically demanding code with respect to full-wave AMR solvers, without compromising cosmological results, with the additional ability to exploit the gas and star physics already implemented in \G, along with its more advanced and exotic beyond-$\Lambda$CDM extensions such as Modified Gravity \citep{Puchwein_Baldi_Springel_2013} or Coupled Dark Energy models \citep[][]{Baldi_etal_2010}.

Given that gravity, as mentioned above, can restore the suppressed power at intermediate scales in the non-linear regime, major observables related to the LSS at such scales may appear similar in both FDM and CDM picture cosmologies at sufficiently low redshifts. For this reason, Lyman-$\alpha$ forest observations could play a crucial role in distinguishing such radically different models of dark matter, being one of the most far reaching direct astrophysical probes in terms of redshift of the LSS observables, sampling the redshift range $z\sim2-5$ \citep[see e.g.][for Lyman-$\alpha$ forest analysis in N-body simulations, with neglected QP dynamical effects]{Irsic17}.

In this paper, we performed several simulations with the main goal of studying the effects of FDM on Lyman-$\alpha$ forest observations in a fully consistent FDM set-up --~i.e. without neglecting the QP during cosmic evolution~-- , in order to constrain the FDM mass. As a by-product of our simulations, we are also able to perform an extended analysis of the statistical and structural properties of haloes, exploiting the large statistical sample at our disposal, to extract valuable information about how FDM affects, among others, the halo mass function as well as the shape and density distribution of dark matter haloes.

The paper is organized as follows: in Section~\ref{sec:theory} we briefly describe the FDM models under consideration, providing all the basic equations that enter our numerical implementation (\ref{sec:fdm}), and review the theoretical background behind Lyman-$\alpha$ forest observations and its physical implications (\ref{sec:theory_la}). In Section~\ref{sec:NM}, we then recall how FDM dynamics is implemented in the \AG code (\ref{sec:AG}), we present the simulation sets performed (\ref{sec:sims}) and the strategy used to extract Lyman-$\alpha$ information (\ref{sec:la_mcmc}); in Sections~\ref{sec:NF} and~\ref{sec:ISHM} the procedures to deal with numerical fragmentation and to match haloes across different simulations are outlined. The results are collected in Section~\ref{sec:results} and presented in decreasing order of interested scale, including the matter power spectrum (\ref{sec:PS}), the Lyman-$\alpha$ statistics (\ref{sec:la_mcmc}) and the structure characterization (\ref{sec:SC}). Finally, in Section~\ref{sec:conclusions} we draw our conclusions.

\section{Theory}
\label{sec:theory}

In this Section we recall the main properties of a light bosonic field in a cosmological framework, how it affects the growth of LSS and how Lyman-$\alpha$ forest analysis can be used to probe these modifications.

\subsection{Fuzzy Dark Matter models}
\label{sec:fdm}

The idea of describing dark matter and its key role in the LSS formation in terms of a ultra-light scalar particles --~i.e. a particle with mass $\sim 10^{-22}eV/c^2$ was introduced in \citet{Hu00}, in which the term \textit{Fuzzy} Dark Matter was used for the first time and the cosmological implications induced by the quantum behaviour of such light dark matter field on linear cosmological perturbations were outlined.

The Schr\"odinger equation describing the dynamics of the bosonic field $\field_i$ associated with a single particle can be written as
\begin{equation}
\label{eq:GPP}
i \hbar \ \de{t} \field_i = - \frac {\hbar^2} {m_{\chi}^2} \nabla^2 \field_i + m_\chi \Phi \field_i
\end{equation}
where $m_\chi$ is the typical mass associated with FDM particles --~often represented in terms of $m_{22}=m_{\chi} 10^{22} c^2 / eV$~-- and $\Phi$ is the gravitational potential, satisfying the usual Poisson equation 
\begin{equation}
\label{eq:poisson}
\nabla^2 \Phi = 4 \pi G a^2 \rho_b \ \delta
\end{equation}
with $\delta=(\rho - \rho_b)/\rho_b$ being the density contrast with respect to the background field density $\rho_b$ \citep{Peebles80}.

Under the assumption that all the particles belong to a Bose-Einstein Condensate, the many-body field $\field$ of a collection of particles factorizes and the collective dynamics follows exactly Eq.~\ref{eq:GPP}. If this is the case, it is possible then to express the many-body field $\field$ in terms of collective fluid quantities as density $\rho$ and velocity $\vec v$, using the Madelung formulation \citep{Madelung27} 
\begin{gather}
\label{eq:madelung}
\vec v = \frac{\hbar}{m_\chi} \Im{ \frac{\vec \nabla \field} {\field} } \\
\rho = m_\chi |\field|^2
\end{gather}
which translates into the usual continuity equation and a modified quantum Navier--Stokes equation reading  
\begin{equation}
\label{eq:NS}
\dot {\vec v} + \left( \vec v \cdot \vec \nabla \right) \vec v =  - \vec \nabla \Phi + \vec \nabla Q 
\end{equation}
where $Q$ is the so-called \textit{Quantum Potential} (QP)
\begin{equation}
\label{eq:QP}
Q = \frac {\hbar^2}{2m_{\chi}^2} \frac{\nabla^2 \sqrt{\rho}}{\sqrt{\rho}}  = \frac {\hbar^2}{2m_{\chi}^2} \left( \frac {\nabla^2 \rho} {2 \rho} - \frac {| \vec \nabla \rho|^2}{4 \rho^2} \right)
\end{equation}
also known as \textit{Quantum Pressure} if expressed in the equivalent tensorial form as
\begin{equation}
\vec \nabla Q = \frac 1 \rho \vec \nabla P_Q %
= \frac {\hbar^2}{2m_{\chi}^2} \ \frac 1 \rho \vec \nabla \left( \frac \rho 4 \vec \nabla \otimes \vec \nabla \ln{\rho} \right).
\end{equation}
The additional QP term accounts for the quantum behaviour of particles with a repulsive net effect that counteracts gravitational collapse below a certain scale, related to the Compton wavelength $\lambda_C = \hbar/m_\chi c$ identified by the boson mass \citep{Hu00}. This can be heuristically viewed as the result coming from two combined effects of quantum wave-like nature: \textit{decoherence}, originating from the Heisenberg uncertainty principle, stirring towards space-filling configurations and \textit{interference} creating oscillatory patterns \citep{Hui16}.

In an expanding universe described by a scale factor $a$ and the derived Hubble parameter $H=\dot a / a$, the linear density perturbation $\delta_k$ in Fourier space satisfies --~in the comoving frame~-- the relation
\begin{equation}
\label{eq:PERT}
\ddot \delta_k + 2 H \dot \delta_k + \left(  \frac{\hbar^2 k^4}{4 m_{\chi}^2 a^4} - \frac{4 \pi G \rho_b } {a^3} \right) \delta_k = 0
\end{equation}
that directly sets the typical scale 
\begin{equation}
\label{eq:kq}
k_Q (a)= \left( \frac{16\pi G \rho_b a^3 m_{\chi}^2}{\hbar^2} \right)^{1/4} a^{1/4} 
\end{equation}
where the gravitational pull is balanced by the QP repulsion, sometimes referred as \textit{quantum Jeans scale} in analogy with the homonym classical one \citep{Chavanis12}.

The growing solution of Eq.~\ref{eq:PERT}, expressed in terms of the dimensionless variable $x(k,a)=\sqrt{6}\ k^2 / k_Q^2(a)$, is
\begin{equation}
\label{eq:growth}
D_+(x) = \left[ \left( 3 - x^2 \right) \cos{x} + 3\ x \sin{x}\right] / x^2
\end{equation}
whose time dependence is bounded from above and below by the large  and small scale limits, respectively, as
\begin{equation}
D_+ \propto \begin{cases}
a & \text{for $k \ll k_Q(a)$} \\
1 & \text{for $k \gg k_Q(a)$} \\
\end{cases}
\end{equation}
thereby recovering the standard $\Lambda$CDM perturbations evolution at large scales and halting growth of small scales over-densities \citep{Marsh16}.

Structures are unable to collapse until the quantum Jeans scale $k_Q(a)$ becomes so little that gravity can overcome the QP repulsive action and, in the linear perturbation regime, will forever carry information about their past suppressed state \citep{Marsh16}. In non-linear simulations, instead, a recovery induced by gravity of the intermediate scales is indeed observed: in terms of matter power spectrum, this implies that a portion of a FDM Universe, observed at a fixed scale, will eventually look like a CDM Universe if a sufficient time for gravity recovery has passed after the crossing of the quantum Jeans scale \citep[as argued also in e.g.][]{Marsh16nl}. 

All this considered, it is clear the reason why FDM models peculiar imprints on LSS are to be looked for at very small scales for low redshifts, while larger scales may provide relevant information only as long as higher redshifts are available to observations. In particular, FDM may reveal its presence in the inner part of small collapsed haloes in the form of a flat solitonic core \citep[see e.g.][]{Schive14,Marsh15CCP,DeMartino18,Lin18} while larger scales may show FDM imprints in the high-redshift gas distribution \citep[][]{Irsic17,Armengaud17,Kobayashi17}.

\subsection{Lyman-\texorpdfstring{$\alpha$}{}
forest}
\label{sec:theory_la}

The \LA forest is the main manifestation of the intergalactic
medium (IGM), the diffuse filamentary matter filling the space between galaxies, and it constitutes a very powerful method for constraining the properties of DM in the small scale ($0.5~{\rm Mpc}/h \lesssim  \lambda \lesssim 20~{\rm Mpc}/h$) and high redshift regime ($2 \lesssim z \lesssim 5$) \citep[see e.g.][]{Viel2005,Viel:2013apy}. The physical observable for \LA experiments is the flux power spectrum $P_{\rm{F}}(k,z)$. Constraints on the matter power spectrum from \LA forest data at small cosmological scales are only limited by the thermal cut-off in the flux power spectrum, introduced by pressure and thermal motions of baryons in the photo-ionised IGM. That is why this astrophysical observable has provided some of the tightest constraints up-to-date on DM scenarios featuring a small-scale power suppression \citep[][]{Irsic:2017ixq,Murgia:2018now}, including FDM models, both in the case where they constitute the entire DM \citep[][]{Irsic17,Armengaud17}, and in the case in which they are a fraction of the total DM amount \citep{Kobayashi17}.

Ultra-light scalar DM candidates are indeed expected to behave differently with respect to standard CDM on scales
of the order of their de Broglie wavelength, where they induce a suppression of the structure
formation, due to their wave-like nature. In particular, for FDM particles with masses $\sim 10^{-22} eV$, such suppression occurs on (sub)galactic scales, being thereby the ideal target for \LA forest observations. Moreover, as we discussed in the previous section, \LA forest observations probe a redshift and scales range in which the difference between $\Lambda$CDM and the FDM models --~for the masses considered~-- is highly significant.

All the limits found in the literature on FDM parameters --~i.e. the mass $m_{\chi}$~-- using \LA observations \citep[as e.g.][]{Irsic17,Armengaud17,Kobayashi17} have been computed by assuming that ultra-light scalars behave as standard pressure-less CDM and by comparing \LA data with flux power spectra obtained from standard SPH cosmological simulations, which completely neglected the QP effects during the non-linear structure evolution. In other words, the non-standard nature of the dark matter candidate was simply encoded in the suppressed initial conditions used as inputs for performing the hydrodynamical simulations.

One of the goals of the present work is to use \AG in order to provide the first fully accurate constraints on the FDM mass, by going beyond the standard dynamical approximation of ignoring the time-integrated QP effect. Including such effect in our numerical simulations is thereby expected to tighten the limits published so far in the literature, since it introduces a  repulsive effect at small scales throughout the simulation evolution that contributes to the matter power spectrum suppression. Besides presenting the new constraints, we will also carry out a meticulous comparison with the bounds determined under the aforementioned approximation, in order to exactly quantify its validity.

\section{Numerical Methods}
\label{sec:NM}

In this Section we briefly review the implementation of the \AG code routines that are devoted to the FDM dynamics \citep[an in depth description featuring analytic and cosmological tests can be found in][]{Nori18}. We then continue presenting how \LA forest observations are modelled and extracted from numerical simulations. Finally, we describe our approach to discriminate spurious haloes --~which are expected to form in particle-based simulations featuring a suppressed power spectrum \citep[see e.g.][]{Wang_White_2007}, such as Warm Dark Matter, Hot Dark Matter, or FDM models~-- from genuine ones, in order to properly take into account the known problem of numerical fragmentation, together with the strategy we used to cross-match haloes in the different simulations.

\subsection{The code: \AG}
\label{sec:AG}

\AG is a module available within the cosmological and hydrodynamical N-Body code \G, a non-public extension of the public {\small GADGET}2 code \citep[][]{Springel05}. It features a new type of particle in the system --~i.e. ultra-light-axion (ULA)~-- whose strongly non-linear quantum dynamics is solved through advanced and refined Smoothed Particle Hydrodynamics (SPH) routines, used to reconstruct the density field from the particle distribution and, therefore, to calculate the QP contribution to particle acceleration.

The general SPH approach relies on the concept that the density field $\rho$ underlying a discrete set of particles can be approximated at particle $i$ position with the weighted sum of the mass $m$ of neighbouring particles $\text{NN}(i)$
\begin{gather}
\rho_i = \sumnn m_j W_{ij},
\end{gather}
where the mass is convolved with a kernel function $W_{ij}$ of choice, characterized by a particle-specific smoothing length $h_i$, and whose extent is fixed imposing
\begin{equation}
\label{eq:NN}
\frac 4 3 \pi h_i^3 \rho_i= \sumnn m_j
\end{equation}
so that only a given mass is enclosed within it.

Once the density field is reconstructed, every observable is locally computed through weighted sums as
\begin{equation}
O_i = \sumnn m_j \frac {O_j} {\rho_j} W_{ij}
\end{equation}
and its derivatives are iteratively obtained with
\begin{equation}
\vec \nabla O_i = \sumnn m_j \frac {O_j} {\rho_j} \vec \nabla W_{ij}
\end{equation}
where the derivation is applied on the window function.

The exact scheme of the SPH algorithm is not fixed, since each observable can be expressed in many analytically equivalent forms that, however, translate into different operative summations. For example, the QP of Eq.~\ref{eq:QP} can be calculated using recursive derivatives of $\rho$,$\sqrt{\rho}$ or $\log{\rho}$ intermediate observables. An important consequence of such flexibility is that different but analytically equivalent expressions will map into operative sums that carry different numerical errors. Among the several strategies that have been employed in the literature to reduce the residual numerical errors \citep[see e.g.][]{Brookshaw85,Cleary99,Colin06}, the following has proven the more stable and accurate for the QP case \citep[see][for a comparison between different implementations]{Nori18}, and will therefore be the one of our choice:

\begin{gather}
\vec \nabla \rho_i = \sumnn m_j \vec \nabla W_{ij} \frac {\rho_j  - \rho_i} {\sqrt{\rho_i \rho_j}} \\
\nabla^2 \rho_i =  \sumnn m_j \nabla^2 W_{ij}  \frac {\rho_j  - \rho_i} {\sqrt{\rho_i \rho_j}} - \frac {|\vec \nabla \rho_i|^2} {\rho_i} \\
\vec \nabla Q_i =  \frac {\hbar^2}{2m_{\chi}^2} \sumnn \frac {m_j} {f_j \rho_j} \vec \nabla W_{ij} 
\left( \frac {\nabla^2 \rho_j} {2 \rho_j} - \frac {| \vec \nabla \rho_j|^2}{4 \rho_j^2} \right).
\end{gather}

\AG has undergone various stability tests and has proven to be not only less numerically intensive with respect to Adaptive Mesh Refinement (AMR) full-wave solvers \citep{GAMER}, due to the intrinsic SPH local approximation, but also to be accurate for cosmologically relevant scales as it agrees both with the linear \citep{axionCAMB} and the non-linear results \citep{Woo09} available in the literature, even if a proper convergence and code comparison test has not yet been performed, since it would be necessary to assess the consistency of different numerical methods at very small scales.

In fact, while cosmological and analytical results --~as e.g. the soliton formation~-- are well recovered by N-Body simulations, interference patterns emerging at very small scales seem more challenging to be represented accurately, due to their oscillatory nature that can be overly smoothed if the resolution --~i.e. the number of particles~-- used is too low. N-body simulations at very high-resolution --~i.e. to the pc level~-- have yet to be performed, but as also argued in more detail in Appendix~\ref{sec:dbb}, it is our opinion that whether interference patterns can be observed or not is ultimately a matter of resolution.

The implementation of FDM physics in \AG includes the possibility to simulate Universes with multiple CDM and FDM species or FDM particles with self- or external interactions, as recently included with the merging of the \AG module with the C-Gadget module of Coupled Dark Matter models \citep{Baldi_etal_2010}.

Moreover, \AG inherits automatically all the large collection of physical implementations --~ranging from gas cooling and star formation routines to Dark Energy and Modified Gravity implementations~-- that have been developed for \G by a wide range of code developers.

All these properties allow to investigate a yet unexplored wide variety of extended FDM models and make of \AG a valuable tool --~complementary to high resolution AMR codes~-- to study the effects of FDM on LSS formation and evolution. In this work, we consider the simplest non-interacting case with the totality of the dark matter fluid composed by FDM.  

\subsection{Simulations}
\label{sec:sims}

In this work, we performed two sets of simulations, for a total number of fourteen cosmological runs. The first set consists in DM-only simulations used to characterize the small scale structures at low redshift --~i.e. down to $z=0$~--, while the second one is evolved to $z=2$ and includes gas particles and a simplified hydrodynamical treatment, as described in Section~\ref{sec:AG}, specifically developed for Lyman-$\alpha$ forest analyses \citep[the so-called "QLYA", or {\em Quick-Lyman-alpha} method, see][]{QLYA}. Both sets consist in three pairs of simulations, one pair for each considered FDM mass, evolved either including or neglecting the effect of the QP in the dynamics --~labelling these two cases as \textit{FDM} and \textit{FDMnoQP}, respectively~--, in order to to assess and quantify the entity of such approximation often employed in the literature. 

Both sets of simulations follow the evolution of $512^3$ dark matter particles in a comoving periodic box with side length of $15 Mpc$, using $1 Kpc$ as gravitational softening. The mass resolution for the dark-matter-only simulations is $2.2124\times10^{6} M_{\odot}$. In all cases we generate initial conditions at $z=99$ using the {\small{2LPTic}} code \citep[][]{2LPTic}, which provides initial conditions for cosmological simulations by displacing particles from a cubic Cartesian grid following a second-order Lagrangian Perturbation Theory based approach, according to a random realisation of the suppressed linear power spectrum as calculated by {\small axionCAMB} \citep{axionCAMB} for the different FDM masses under investigation. To ensure a coherent comparison between simulations, we used the same random phases to set up the initial conditions. In particular, the FDM masses $m_{\chi}$ considered here are $2.5\times 10^{-22}$, $5\times 10^{-22}$ and $2.5\times 10^{-21} \text{eV}/c^2$, in order to sample the  mass range preferred by the first \LA constraints in the literature \citep[see in particular][]{Irsic17,Armengaud17,Kobayashi17}, obtained through N-body simulations with approximated dynamics.

Cosmological parameter used are $\Omega_{\rm m} = 0.317$, $\Omega_{\Lambda} = 0.683$, $\Omega_{\rm b} = 0.0492$ and $H_0 = 67.27$ km/s/Mpc, $A_s = 2.20652\times10^{-9}$ and $n_s = 0.9645$. A summary of the simulation specifications can be found in Tab.~\ref{tab:SIMS}.

\begin{table*}
\caption{Summary of the properties of the simulations set used for structure characterization.}
\label{tab:SIMS}
\begin{tabular}{cccccc}
\hline
Model    & QP in dynamics & $m_{\chi} \ [10^{-22}\text{eV}/c^2]$ & N haloes & N genuine haloes & $M_{\text{cut}}\ [10^{10} M_{\odot}]$ \\
\hline
LCDM    & $\times$       & -                            & 57666       & 56842               & -                \\
\hline
FDM-25  & $\checkmark$   & 25                            & 25051       & 13387               & $0.04056$                 \\
FDM-5   & $\checkmark$   & 5                            & 10058       & 2736                & $0.1645$                 \\
FDM-2.5 & $\checkmark$   & 2.5                          & 8504        & 1301                & $0.3151$                 \\
\hline
FDMnoQP-25  & $\times$       & 25                            & 25432       & 13571               & $0.04056$                 \\
FDMnoQP-5   & $\times$       & 5                            & 10376       & 2856                & $0.1645$                 \\
FDMnoQP-2.5 & $\times$       & 2.5                          & 8819        & 1374                & $0.3151$                 \\
\hline
\end{tabular}
\end{table*}

\subsection{Lyman-\texorpdfstring{$\alpha$}{}
forest}
\label{sec:la_mcmc}

The flux power spectrum $P_F(k,z)$ is affected both by astrophysical and cosmological parameters. It is therefore crucial to accurately quantify their impact in any investigation involving the flux power as a cosmological observable. To this end, our analysis is based on a set of full hydrodynamical simulations which provide a reliable 
template of mock flux power spectra to be compared with observations.

For the variations of the mean \LA forest flux, $\bar{F}(z)$, we have explored models up to 20\% different than the mean evolution given by \cite{Viel:2013apy}.

We have varied the thermal history of the Intergalactic Medium (IGM) in the form of the
amplitude $T_0$ and the slope ${\gamma}$ of its temperature-density
relation, generally parameterized as $T=T_0(1+\delta)^{{\gamma}-1}$, with $\delta$ being the IGM over-density \citep[][]{hui97}.
We have then considered a set of three different temperatures at mean density, $T_0(z = 4.2) = 7200, 11000, 14800$~K, which
evolve with redshift, as well as a set of three values for the slope of the temperature-density relation, ${\gamma}(z = 4.2) = 1.0, 1.3, 1.5$. The reference thermal history has been chosen to be defined by $T_0(z = 4.2) = 11000$ and ${\gamma}(z = 4.2) = 1.5$, providing a good fit to observations \citep[][]{bolton17}. Following the conservative approach of~\cite{Irsic17}, we have modelled the redshift evolution of ${\gamma}$ as a power law, such that ${\gamma}(z) = {\gamma}^A[(1+z)/(1+z_p)]^{{\gamma}^S}$, where the pivot redshift $z_p$ is the redshift at which most of the Lyman-$\alpha$ forest pixels are coming from (i.e.~$z_p = 4.2$ for MIKE/HIRES+XQ-100). However, in order to be agnostic about the thermal history evolution, we let the amplitude $T_0(z)$ free to vary in each redshift bin, only forbidding differences greater than 5000~K between adjacent bins \citep[][]{Irsic:2017ixq}.

Furthermore, we have also explored several values for the cosmological parameters $\sigma_8$,~i.e.~the normalisation of the matter power spectrum, and $n_{\rm eff}$,~namely the slope of the matter power spectrum at the scale of \LA forest (0.009 s/km), in order to account for the effect on the matter power spectrum due to changes in its initial slope and amplitude \citep[][]{seljak2006,McDonald:2004eu,Arinyo-i-Prats:2015vqa}. We have therefore considered five different values for $\sigma_8$ (in the interval $[0.754, 0.904]$) and $n_{\rm eff}$ (in the range $[-2.3474, -2.2674]$).

We have also varied the re-ionization redshift $z_{\rm rei}$, for which we have considered the three different values $z_{\rm rei} = 7,9,15$, with $z_{\rm rei} = 9$ being the reference value and, finally, we have considered ultraviolet (UV) fluctuations of the ionizing background, that may have non-negligible effects at high redshift. The amplitude of this phenomenon is parameterized by the parameter $f_{\rm UV}$: the corresponding template is built from a set of three models with $f_{\rm UV} = 0, 0.5, 1$, where
$f_{\rm UV} = 0$ is associated with a spatially uniform UV background.

Based on the aforementioned grid of simulations, we have performed a linear interpolation between the grid points in such multidimensional parameter space, to obtain predictions of flux power for the desired models.

We have to note that the thermal history implementation of \citet{Irsic17} and the one used in this work are slightly different. For this reason, since the simulations of the grid were performed without the introduction of the QP in the dynamics, we mapped our results into the grid ones using the ratio between FDM and FDMnoQP simulations. This is, of course, not an exact procedure but we assume that the ratio of flux power spectrum with and without quantum pressure is relatively insensitive to the thermal history \citep{Murgia:2018now}.

In order to constrain the various parameters we have used a data set given by the combination of intermediate and high resolution \LA forest data from the XQ-$100$ and the HIRES/MIKE samples of QSO spectra, respectively.
The XQ-$100$ data are constituted by a sample of medium resolution and intermediate signal-to-noise QSO spectra, obtained by the XQ-$100$ survey, with emission redshifts $3.5 \leq z \leq 4.5$ \citep[][]{xq100}. The spectral resolution of the X-shooter spectrograph is $30-50 km/s$,
depending on the wavelength. The flux power spectrum $P_{\rm F}(k,z)$
has been calculated for a total of $133$ $(k,z)$ data points in the ranges $z=3,3.2,3.4,3.6,3.8,4,4.2$ and $19$ bins in $k$-space in the
range $0.003-0.057 s/km$ \citep[see][for a more detailed description]{Irsic:2017sop}.
MIKE/HIRES data are instead obtained with the HIRES/KECK and the MIKE/Magellan spectrographs, at redshift bins
$z=4.2,4.6,5.0,5.4$ and in $10$ $k$-bins in the interval $0.001-0.08 s/km$, with spectral resolution of $13.6$ and $6.7 km/s$, for HIRES and MIKE, respectively \citep[][]{Viel:2013apy}. As in the analyses by \citet{Viel:2013apy} and \citet{Irsic:2017ixq}, we have imposed a conservative cut on the flux power spectra obtained from MIKE/HIRES data, and only the measurements with $k > 0.005 s/km$ have been used, in order to avoid possible systematic uncertainties on large scales due to continuum fitting. Furthermore, we do not consider the highest redshift bin for MIKE data, for which the error bars on the flux power spectra are very large \citep[see][for more details]{Viel:2013apy}.
We have thus used a total of of $182$ $(k,z)$ data points.
Parameter constraints are finally obtained with a Monte Carlo Markov Chain (MCMC) sampler which samples the likelihood space until convergence is reached.

\subsection{Numerical Fragmentation}
\label{sec:NF}

For cosmological models whose LSS properties depart sensibly from $\Lambda$CDM only at small scales --~as FDM models~--~, the thorough analysis of the statistical overall properties and the specific inner structures of haloes represents the most relevant and often largely unexploited source of information.In N-body simulations, this implies the use of a suitable clustering algorithm to build a halo catalogue in order to identify gravitationally bound structures that can then be studied in their inner structural properties.

In this work, we rely on the {\small{SUBFIND}} routine already implemented in \G, a two step halo-finder which combines a Friends-Of-Friends (FoF) algorithm \citep[][]{Davis_etal_1985} to find particle clusters --~that defines the primary structures of our halo sample~-- with an unbinding procedure to identify gravitationally bound substructures within the primary haloes \citep{Springel_etal_2001}. Hereafter, we use the terms \textit{primary structures} to identify the substructures of each FoF group containing the  most gravitationally bound particle, \textit{subhaloes} for the non-primary structures and \textit{haloes} when we generally consider the whole collection of structures found.

However, a long-standing problem that affects N-body simulations, when characterized by a sharp and resolved cut-off of the matter power spectrum, has to be taken into account in the process of building a reliable halo sample. This is the so-called \textit{numerical fragmentation}, i.e. the formation of artificial small-mass spurious clumps within filaments \citep[see e.g.][]{Wang_White_2007, Schneider_etal_2012, Lovell13, Angulo_Hahn_Abel_2013, Schive16}.

While it has been initially debated whether the nature of such fragmentation was to be considered physical or numerical, the detailed analysis by \citet{Wang_White_2007} showed that in Warm and Hot Dark Matter simulations \citep[as e.g.][]{Bode00} --~which are characterised by a highly suppressed matter power spectrum~-- the formation of small mass subhaloes was resolution dependent and related to the large difference between force resolution and mean particle separation \citep[as already suggested by][]{Melott_Shandarin_1989}.

To identify spurious haloes in simulations and select a clean sample to study and characterize the structures of FDM haloes in each simulation, we take cue from the procedure outlined in \citet{Lovell13}: in particular, we use the mass at low redshift and the spatial distribution of particles as traced back in the initial conditions as proxies for the artificial nature of haloes as described below.

In fact, the more the initial power spectrum is suppressed at small scales, the more neighbouring particles are coherently homogeneously distributed, thus facilitating the onset of artificially bounded and small ensembles that eventually outnumber the physical ones. As already shown by \citet{Wang_White_2007}, the dimensionless power spectrum peak scale $k_{peak}$ and the resolution of the simulation --~i.e. described through the mean inter-particle distance $d$~-- can be related together to get the empirical estimate
\begin{equation}
\label{eq:MLIM}
M_{lim} = 10.1\ \rho_b\ d\ /\ k^2_{peak}
\end{equation}
describing the mass at which most of the haloes have a numerical rather than a physical origin. In \citet{Lovell13}, this mass is used as a pivotal value for the mass $M_{CUT}$ used to discriminate genuine and spurious haloes --~lying above and below such threshold, respectively~-- which is set as $M_{CUT} = 0.5 M_{lim}$.

In addition to the mass discriminating criterion, \citet{Lovell13} showed that particles that generate spurious haloes belong to degenerate regions in the initial conditions and are more likely to lie within filaments, stating that the reconstructed shape of the halo particles ensemble in the initial conditions can be used to identify spurious structures. N-Body initial conditions are generally designed as regularly distributed particles on a grid from which are displaced in order to match the desired initial power spectrum. Hence, numerical fragmentation originates mostly from particles lying in small planar configurations, belonging to the same row/column domain or a few adjacent ones.

Therefore, we need a method to quantitatively describe the shape of subhaloes and of the distribution of their member particles once traced back to the initial conditions of the simulation. To this end, we resort to the inertia tensor of the particle ensemble
\begin{equation}
I_{ij} = \sum_{particles} m\ (\hat e_i \cdot \hat e_j)\ |r|^2 - (\vec r \cdot \hat e_i)\ (\vec r \cdot \hat e_j)
\end{equation}
where $m$ and $\vec r$ are the particle mass and position respectively and $\hat e$ are the unit vectors of the reference orthonormal base. The eigenvalues and the eigenvectors of the inertia tensor represent the square moduli and unitary vectors of the three axes of the equivalent triaxial ellipsoid with uniform mass distribution. We define $a \geq b \geq c$ the moduli of the three axes and the sphericity $s = c/a$ as the ratio between the minor and the major ones: a very low sphericity will characterize the typical degenerate domains of numerical fragmentation. 

For these reasons, we use the combined information carried by the mass and the sphericity in the initial condition to clean the halo catalogues from spurious ones by applying independent cuts on both quantities as will be detailed below.

\begin{figure*}
\centering
\begin{tabular}{cc}
$z=0$ & $z=99$ \\
\\
\includegraphics[width=0.48\textwidth,trim={0.3cm 1.3cm 1.1cm 2.7cm},clip]{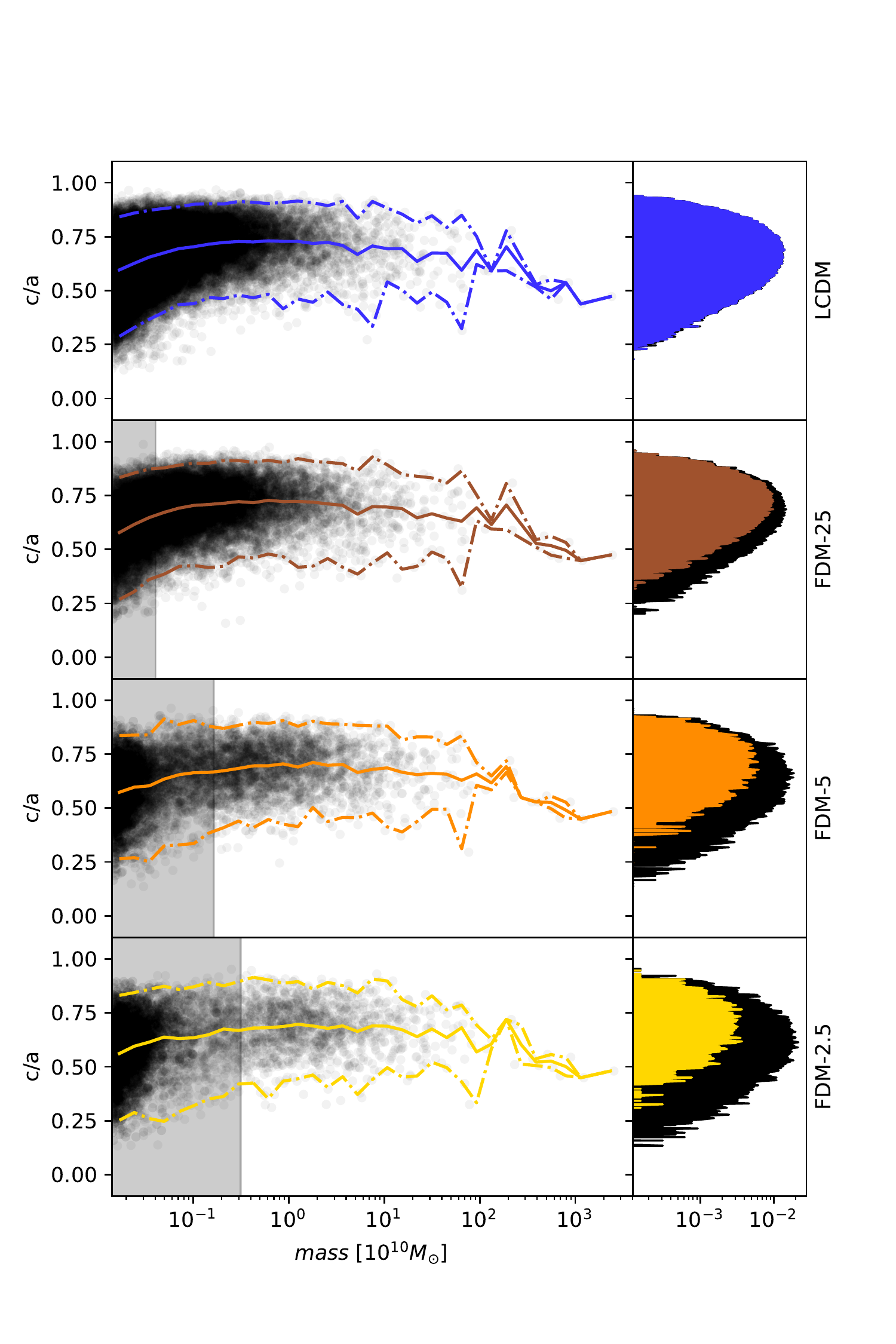} &
\includegraphics[width=0.48\textwidth,trim={0.3cm 1.3cm 1.1cm 2.7cm},clip]{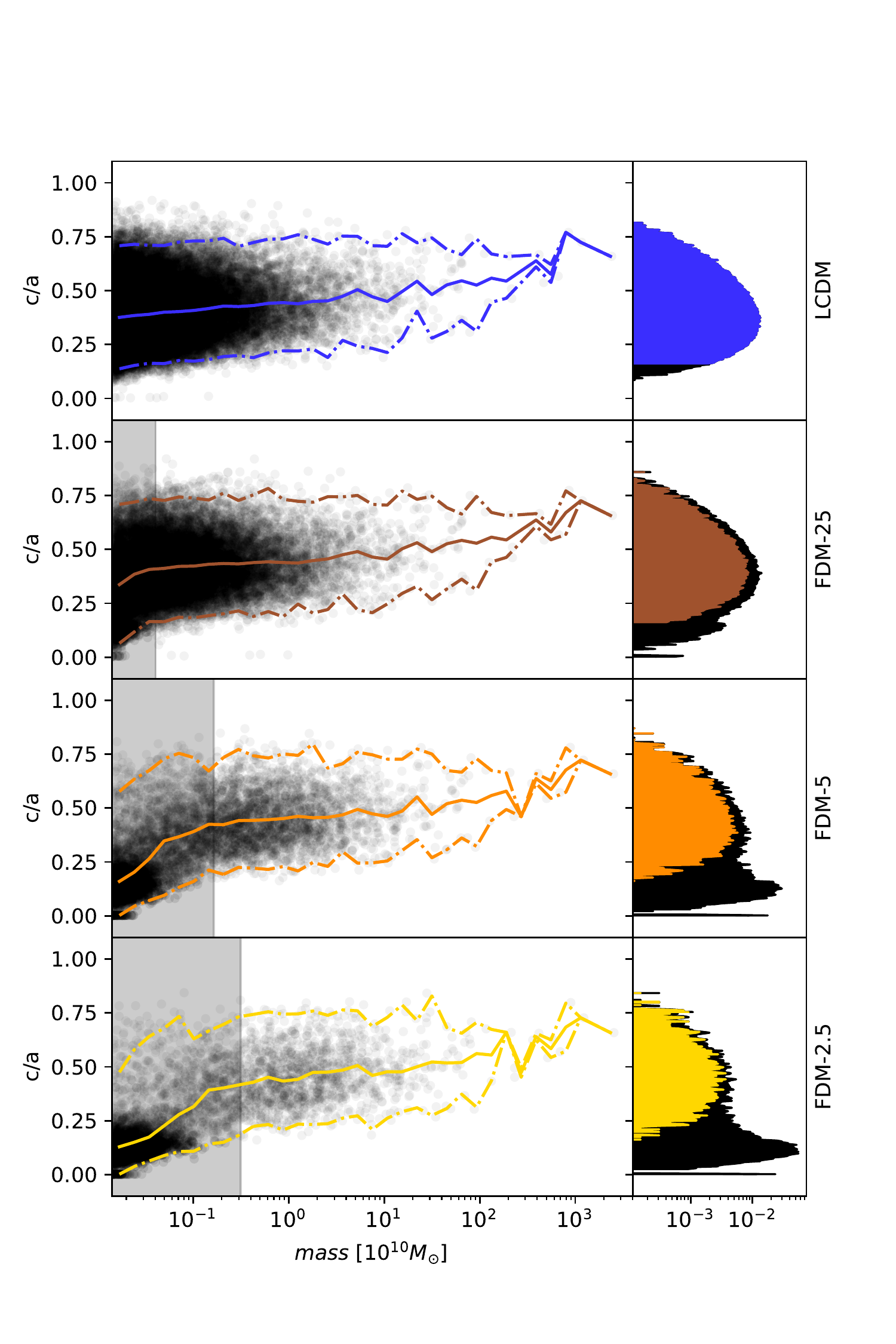} \\
\includegraphics[width=0.48\textwidth,trim={0.3cm 0.2cm 1.1cm 1.4cm},clip]{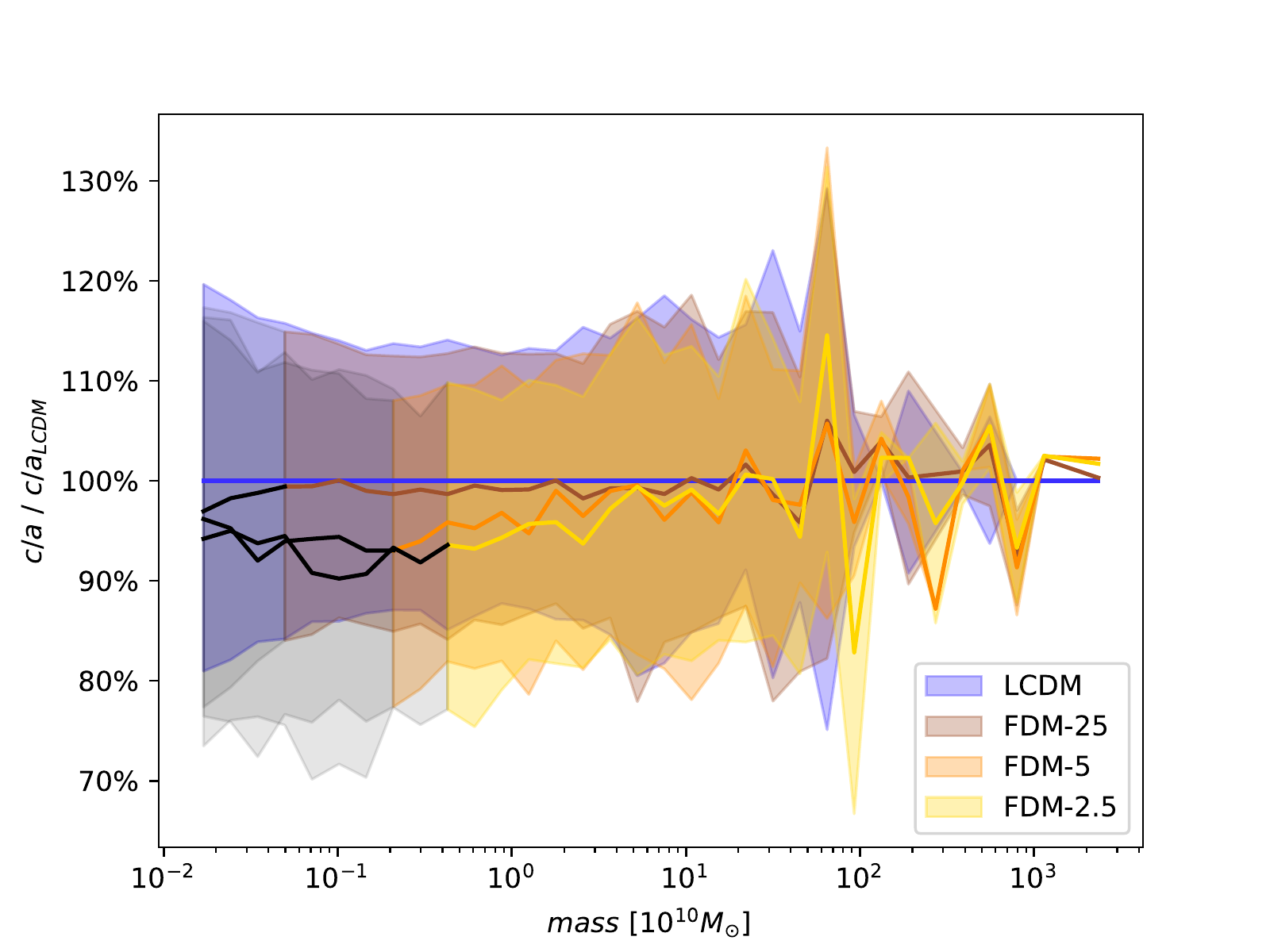} &
\includegraphics[width=0.48\textwidth,trim={0.3cm 0.2cm 1.1cm 1.4cm},clip]{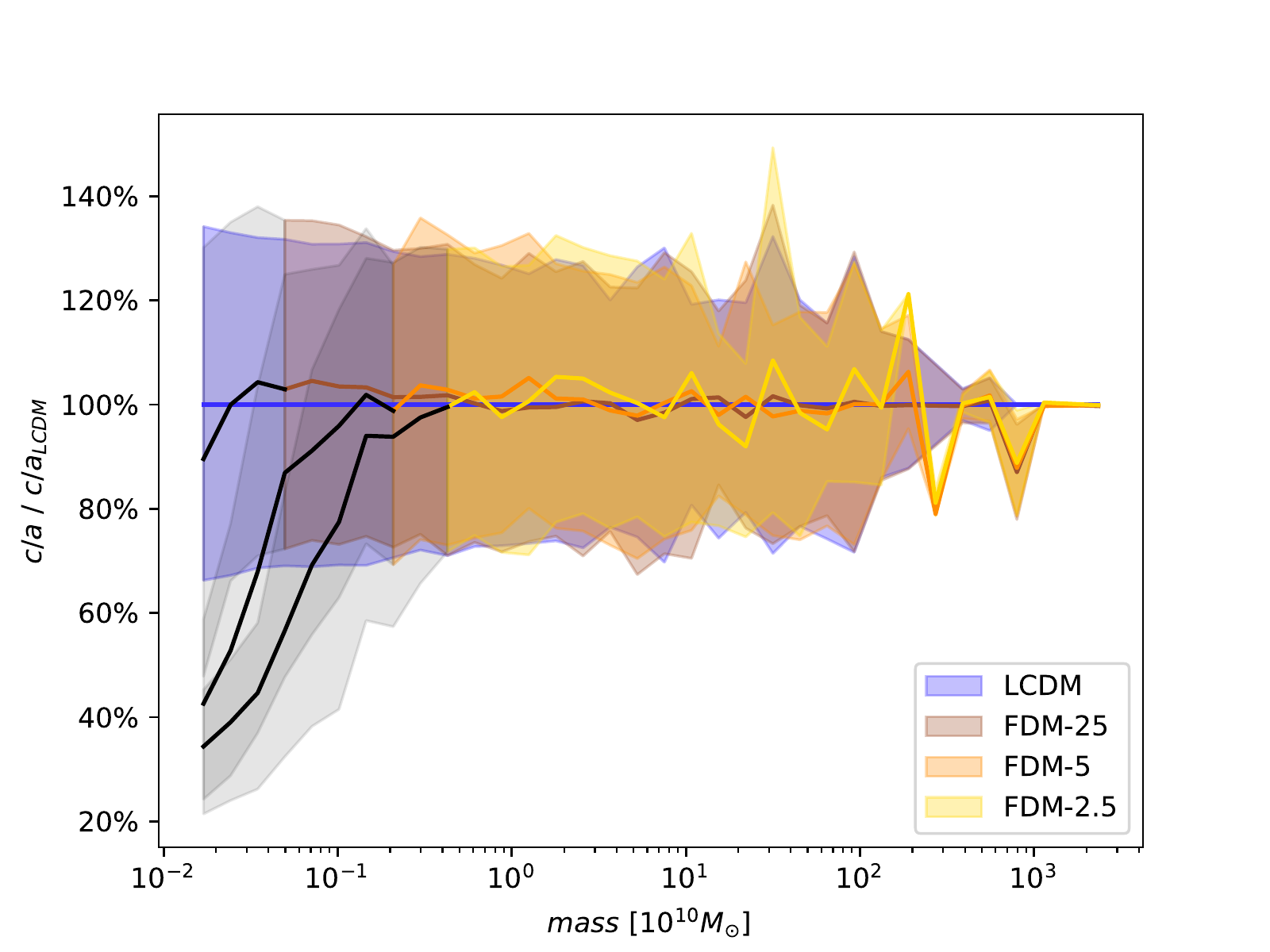} \\
\end{tabular}
\caption{Sphericities of all dark matter particles ensembles as found by {\small{SUBFIND}} as function of their mass ({\it upper panels}) at redshifts $z=0$ and $z=99$ (\textit{left} and \textit{right panels}, respectively). The black-shaded area represents the discarded region below the different mass cuts $M_{CUT}$, corresponding to each model. Each black dot represents a subhalo and the solid (dot-dashed) lines describe the median (99th percentile) of the total distribution, which are all gathered and contrasted with $\Lambda$CDM in the lower panel. The total sphericity distribution --~integrated in mass~-- is represented in the side panels where the contribution of the discarded sample to medians and distributions are portrayed in black. \textit{Lower panels} feature the median of the mass-sphericity distributions, presented as the ratio with respect to $\Lambda$CDM. The shaded areas, corresponding to the $\pm 1 \sigma$ of the distribution, are colour-coded as in the upper panels. The blackened median and shaded areas represent the excluded portion of the sphericity distributions below the corresponding $M_{CUT}$.}
\label{fig:SPHERICITIES}
\end{figure*}

In Fig.~\ref{fig:SPHERICITIES} the mass-sphericity distributions of the different simulations are plotted at $z=0$ (\textit{upper left panel}) and at $z=99$ (\textit{upper right panel}) where each point represents a halo identified by {\small{SUBFIND}}, without applying any selection. Solid and dash-dotted lines denote the median and the $99th$ percentile of the distribution; in the side panels we display the cumulative distributions, where the contribution of spurious haloes is highlighted in black.

By looking at the two panels, it is possible to notice that the total cumulative sphericity distribution at low redshift is fairly model independent, so that distinguishing spurious haloes from genuine ones is impossible. However, if we trace the particle ensembles of each halo found at $z=0$ back to the initial conditions at redshift $z=99$, using particles ID, and we study the resulting reconstructed mass-sphericity relation, the anomalous component of the distribution associated with spurious haloes clearly emerges as a low-sphericity peak, which is more pronounced for smaller values of the FDM particle mass.

In fact, as the mass $m_{\chi}$ decreases, the smoothing action of the QP becomes more efficient, inducing homogeneity at larger and larger scales in the initial conditions and increasing, consequently, the contamination of numerical fragmentation. It clearly appears that the population of haloes in the initial conditions is homogeneously distributed in $\Lambda$CDM while a bimodal structure emerges at lower and lower FDM mass. In particular, an increasing number of haloes are located in a small 
region characterized by low mass ($M \lesssim 10^{9} M_{\odot}$) and low sphericity ($s \lesssim 0.20$).

As there is no theoretical reason why the QP should favour the collapse of ensembles with very low sphericities in the initial conditions with respect to the $\Lambda$CDM case, we consider this second population as the result of numerical fragmentation. 

As in \citet{Lovell13}, we choose to compute $M_{CUT} = 0.5 M_{lim}$ using Eq.~\ref{eq:MLIM} --~one $M_{CUT}$ for each value of the FDM mass, as reported in Tab.~\ref{tab:SIMS}~--, that define the upper bound of the discarded mass regions, i.e. the black-shaded areas in all panels of Fig.~\ref{fig:SPHERICITIES}.  %that for our simulations correspond to $8.11 \times 10^{8} M_{\odot}$, $6.30\times 10^{9} M_{\odot}$ and $3.29\times 10^{9} M_{\odot}$ for $m_{22}$ equal to $25$, $5$ and $2.5$, respectively.

It is interesting to notice that the masses $M_{CUT}$ appear to be very close to the values at which the sphericity medians of the simulation sample --~in the initial conditions~-- depart from the ones of $\Lambda$CDM, as can be seen in the \textit{lower-right} panel of Fig.~\ref{fig:SPHERICITIES}. As the $M_{CUT}$ values we obtain are slightly larger compared to these departing values, we confirm the choice of the former over the latter, as a more conservative option for the mass thresholds dividing spurious from genuine haloes.

In \citet{Lovell13}, the selection in terms of initial sphericity was operationally performed discarding every halo with sphericity lower than $s_{CUT}=0.16$, equal to the $99th$ percentile of the distribution of haloes with more than $100$ particles in the $\Lambda$CDM simulation. In our set of simulations, a similar value denotes the $99th$ percentile as measured at the $M_{CUT}$ mass in each simulations, so we adopt it as our own threshold in sphericity. Let us stress that the haloes that are discarded through sphericity selection in the initial conditions have sphericities at $z=0$ that are statistically consistent with the genuine sample, making their numerical origin impossible to notice based only on the sphericity distribution at $z=0$. However, the mass constraint is far more rigid than the sphericity one in all models but $\Lambda$CDM, where no mass limit is imposed. 

Finally, in Tab.~\ref{tab:FDMnoQPhaloes} we have summarized the comparison of the number of haloes in the FDMnoQP set-up with respect to the corresponding FDM set-up, presented as the ratio of the total number of haloes found by {\small{SUBFIND}} and the number of genuine haloes remaining after the exclusion of spurious ones. It is possible to see that in the FDMnoQP simulations, for the three FDM masses considered, the total number of haloes is overestimated by a factor $\sim 2.5\%$ on average while the genuine haloes excess becomes more important as the FDM mass decreases, up to $5.6 \%$ for $m_{22}=2.5$. This means that neglecting the effects of the QP during the simulation leads to the formation of haloes which are not present when the full QP dynamics is taken into account and that, using our \textit{\`{a} la} \citet{Lovell13} spurious detection selection, such haloes pass the numerical fragmentation test and contaminate any halo statistical property characterization.

\begin{table}
\caption{The total and genuine number of haloes, presented as the ratio between the simulations neglecting and considering the QP dynamical effects.}
\label{tab:FDMnoQPhaloes}
\centering
\begin{tabular}{ccc}
\hline
$m_{\chi}\ [10^{-22}\text{eV}/c^2]$ & N haloes & N genuine haloes \\
\hline
25                                  & 101.6\%  & 101.4\%             \\
5                                   & 103.5\%  & 104.4\%             \\
2.5                                 & 103.1\%  & 105.6\%             \\
\hline
\end{tabular}
\end{table}

\subsection{Inter-simulations halo matching}
\label{sec:ISHM}

In FDM models, as we said in the previous sections, not only the initial power spectrum of matter perturbation is suppressed at small scales, thereby preventing the formation of small mass structures, but the dynamical evolution of density perturbations changes due to the effect of the QP, intimately affecting the development of structures during the whole cosmological evolution by opposing gravitational collapse. The implementation of such effect in \AG breaks the one-to-one correspondence of the spatial position of collapsed structures in simulations with different FDM masses --~especially for smaller objects~--, despite the identical random phases used to set up the initial conditions. 

We indeed expect bigger haloes not to change dramatically their position at low redshift across different simulation, while this is not the case for lighter subhaloes which are more affected by the evolving local non-linear balance between gravity and the QP of the environment.

This makes it more difficult to identify matching collapsed objects of common origin across the simulations, to study how FDM models affect the inner structure of haloes on a halo-to-halo basis.

To this end, we devise an iterative matching procedure, to be repeated until no more couples are found, as the following: given a halo $i$ at position $\vec r_i$ and total mass $m_i$ in simulation $A$,
\begin{enumerate}
\item select all haloes $j$ belonging to simulation $B$ as potential counterparts if $| \vec r_i - \vec r_j | / (a_i + a_j) < \tilde{R}$ where $a_i$ and $a_{j}$ are the major axes of the haloes computed through the inertia tensor of all their member particles. 
\item within the ensemble selected at the previous point, retain only the haloes ${k} \subseteq {j}$ whose masses satisfy the condition $| m_i - m_k | / (m_i + m_k) < \tilde{M}$ 
\item if more than one halo ${l} \subseteq {k}$ is left, then choose the one for which $| \vec r_i - \vec r_l | / (a_i + a_l)$ is minimum.
\item after having considered all the haloes in $A$, if more than one are linked to the same halo $l$ belonging to $B$, choose the couple $(i,l)$ that minimizes $\left[| \vec r_i - \vec r_l | / (a_i + a_l) \right]^2 + \left[| m_i - m_l | / (m_i + m_l) \right]^2$, in order give the same weight to the two criteria.
\end{enumerate}

This method is flexible enough to account for the shift in mass and position we expect from simulations with different FDM mass models, but conservative enough to ensure the common origin of the subhalo couples. Moreover, using the combination of position and mass filters, we are able to discriminate couples in all mass ranges: position filtering is weaker constraint in the case of bigger haloes --~since they occupy a big portion of a simulation~-- where instead the mass filter is very strict; vice-versa, it is more powerful for smaller haloes for which the mass filter select a large number of candidates.

Operatively, we use the previous procedure to match haloes in each simulation with the $\Lambda$CDM one and we refer to the subset of haloes that share the same $\Lambda$CDM companion across all the simulations as the \textit{common} sample.

For geometrical reasons, we set the limit value for $\tilde{R}$ to be $0.5$: this represents the case in which two haloes with the same major axis $a$ have centres separated exactly by the same amount $a$. The configurations that are selected by point $(i)$ are the ones for which the distance between the halo centres is less or equal the smallest major axis between the two. A higher value for $\tilde{R}$ would include genuine small haloes that have been more subject to dynamical QP drifting but would also result in a spurious match of bigger haloes. For these reasons, we adopt $\tilde{R}=0.5$, checking that the selected sample gains or loses $\sim 5\%$ of components if values $0.45$ and $0.55$ are used, without modifying the overall statistical properties of the sample itself.

\begin{table}
\caption{Number of common matches across LCDM and FDMs simulations, using different values of the parameter $\tilde{M}$ representing the minimum allowed ratio between the minimum and maximum masses of each candidate couple.}
\label{tab:MCUT}
\centering
\begin{tabular}{ccc}
\hline
$\tilde{M}$ & $m_{min} / m_{max}$ & N matches \\
\hline
1/39              & 95\%                & 53        \\
1/19              & 90\%                & 162       \\
3/37              & 85\%                & 234       \\
1/9               & 80\%                & 279       \\
1/7               & 75\%                & 304       \\
1/3               & 50\%                & 346       \\
3/5               & 25\%                & 361       \\
1                 & 0\%                 & 389       \\
\hline
\end{tabular}
\end{table}

With respect to $\tilde{M}$ at point $(ii)$, instead, we applied the matching algorithm using several values, each denoting a specific threshold of the minimal value allowed for the mass ratio of halo couples in order to be consider as a match. As reported in Tab.~\ref{tab:MCUT}, more than $60\%$ of all the matching haloes across LCDM and FDMs simulations without mass selection --~$\tilde{M} = 1$ case~-- have a mass ratio in the $100-85\%$ ratio range and almost $80\%$ in the $100-75\%$ range. In order not to spoil our matching catalogue, especially with very close but highly different in mass halo couples, we choose the limiting value of $\tilde{M} = 1/7$.

\bigskip

\section{Results}
\label{sec:results}

In this Section we present the results obtained from our simulations in decreasing order of scale involved, starting from the matter power spectrum, to the simulated Lyman-$\alpha$ forest observations, to the statistical characterization of halo properties and their density profiles.

\subsection{Matter Power spectrum}
\label{sec:PS}

\begin{figure*}
\begin{tabular}{cc}
$z=9$ & $z=5.4$ \\
\includegraphics[width=0.48\textwidth,trim={0.1cm 0.1cm 1.1cm 1.3cm},clip]{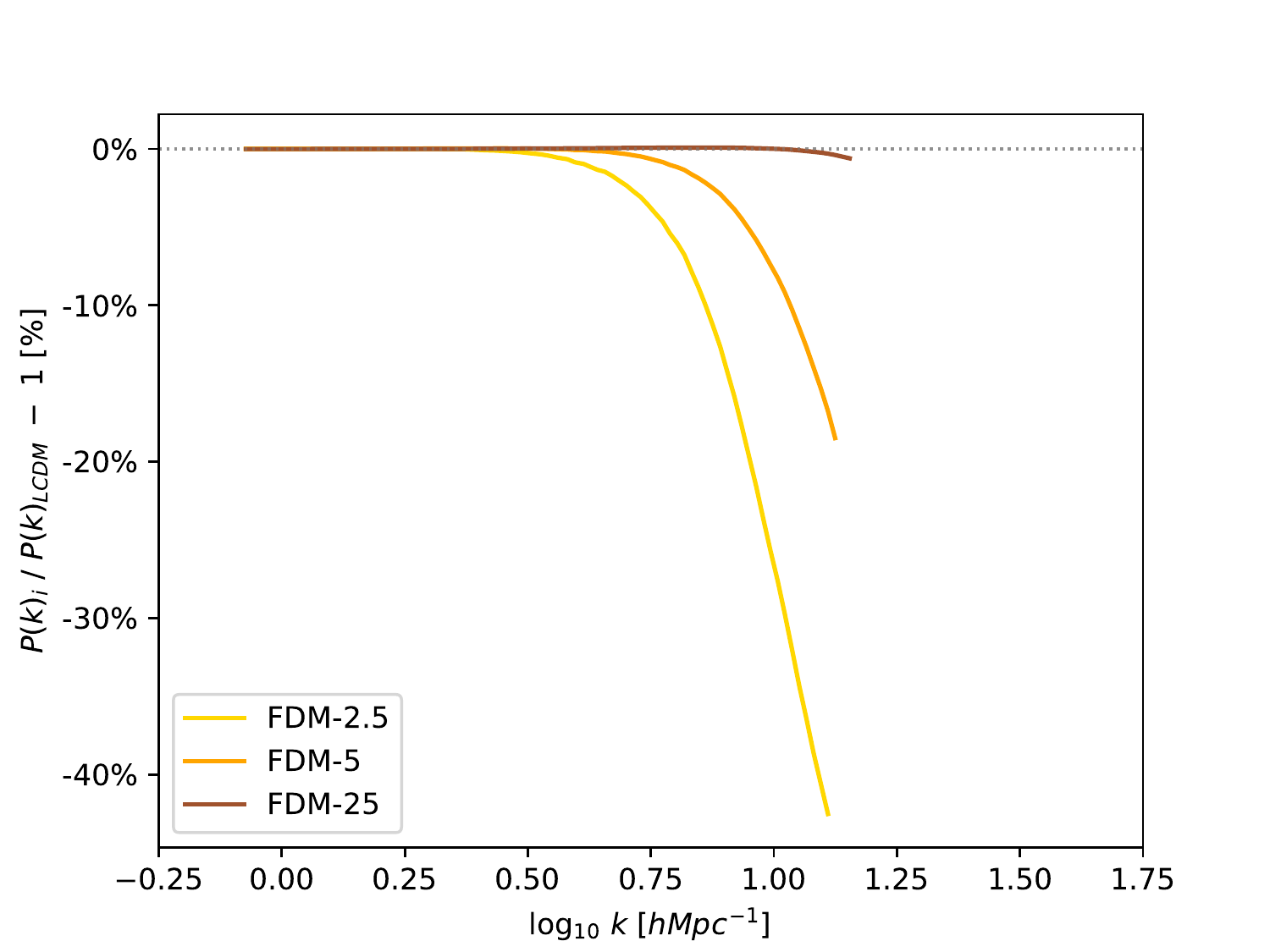}&
\includegraphics[width=0.48\textwidth,trim={0.1cm 0.1cm 1.1cm 1.3cm},clip]{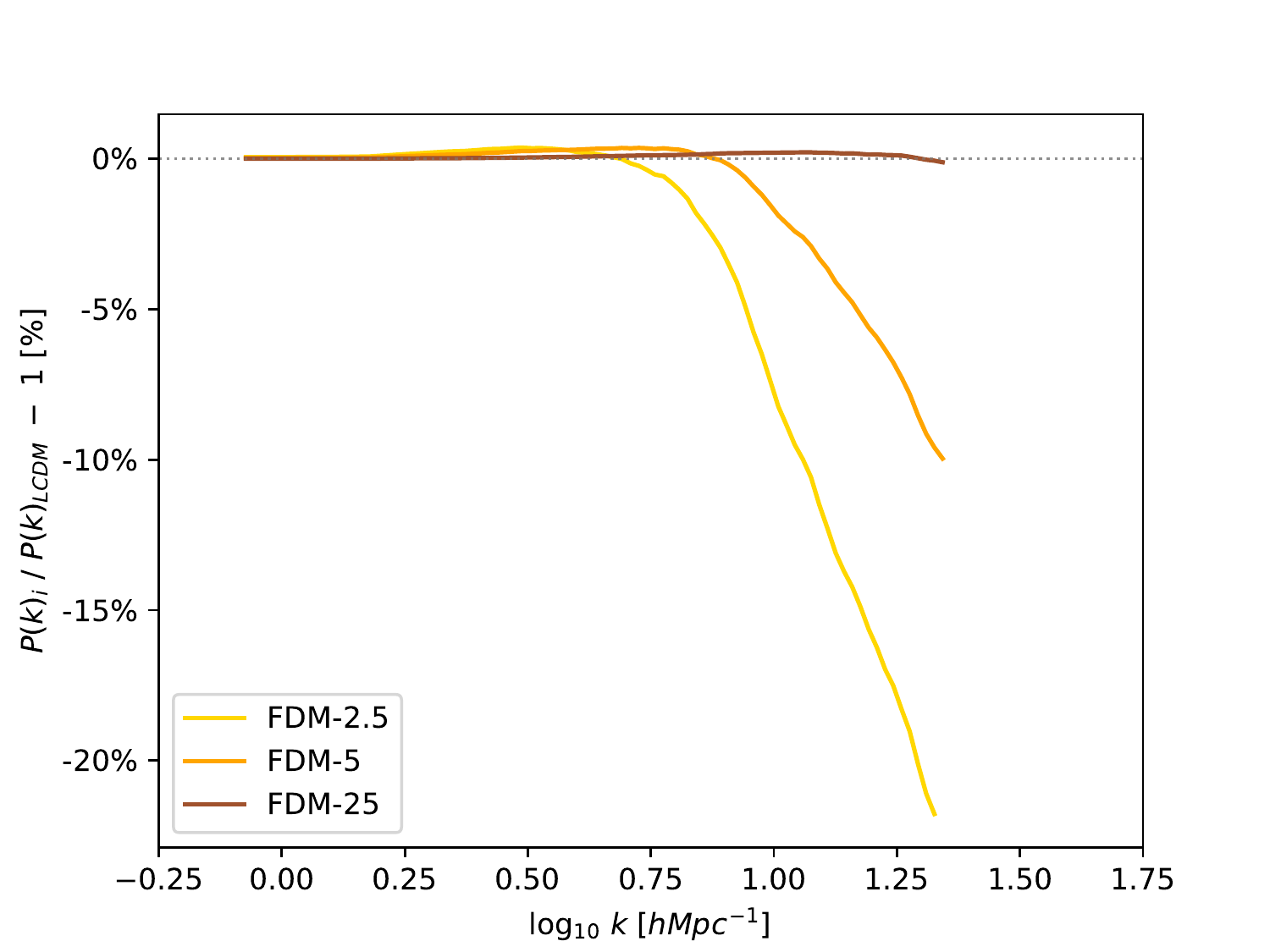}\\
\\
$z=3.6$ & $z=1.8$ \\
\includegraphics[width=0.48\textwidth,trim={0.1cm 0.1cm 1.1cm 1.3cm},clip]{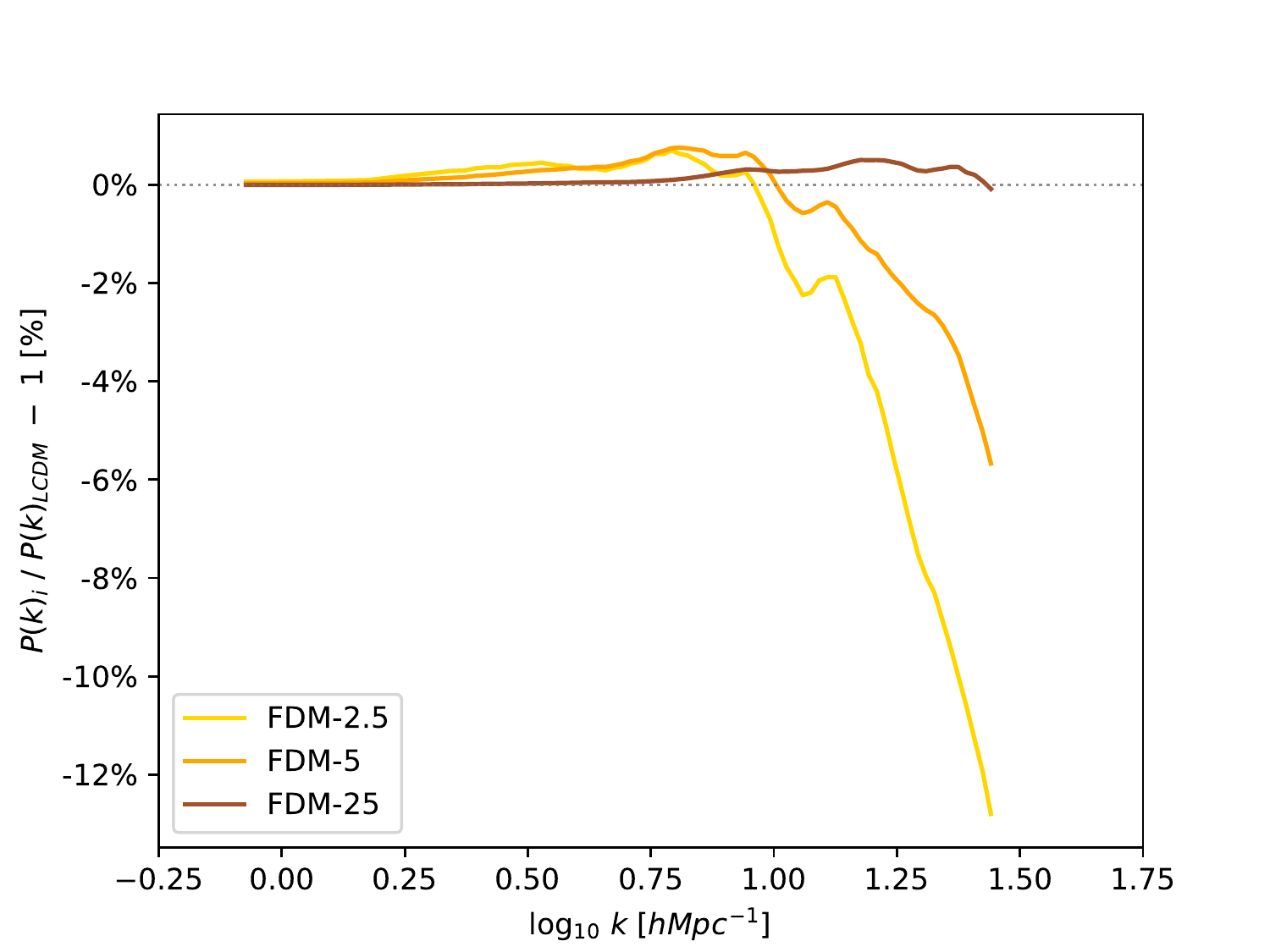}&
\includegraphics[width=0.48\textwidth,trim={0.1cm 0.1cm 1.1cm 1.3cm},clip]{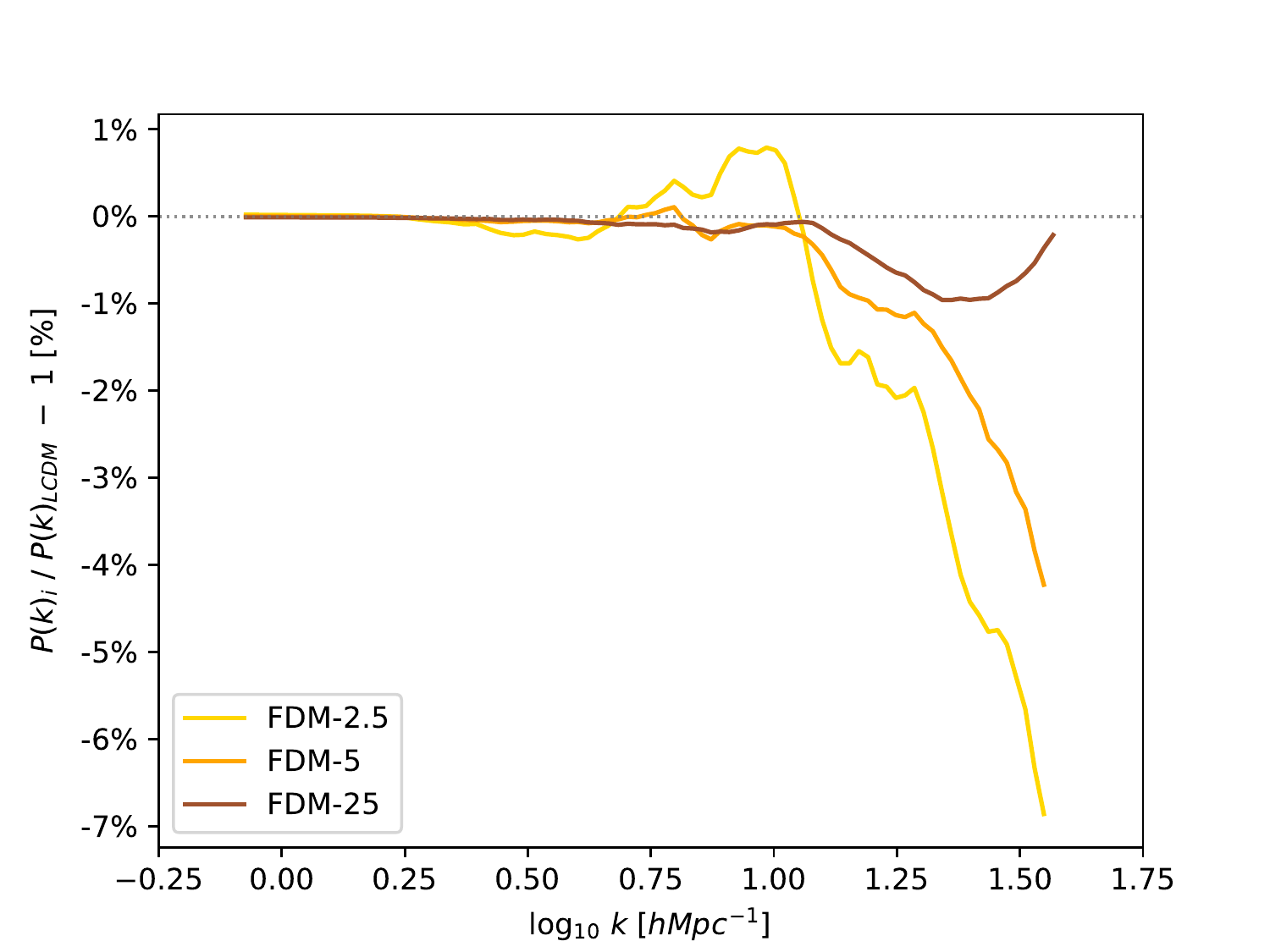}\\
\end{tabular}
\caption{Matter power spectrum of FDM models contrasted with LCDM at different redshifts.}
\label{fig:PS_FDMvsLCDM}
\end{figure*}

The relative difference of the matter power spectrum of the various FDM models with respect to $\Lambda$CDM  at four different redshifts is displayed in Fig.~\ref{fig:PS_FDMvsLCDM}. 

\begin{figure*}
\begin{tabular}{cc}
$z=9$ & $z=5.4$ \\
\includegraphics[width=0.48\textwidth,trim={0.1cm 0.1cm 1.1cm 1.3cm},clip]{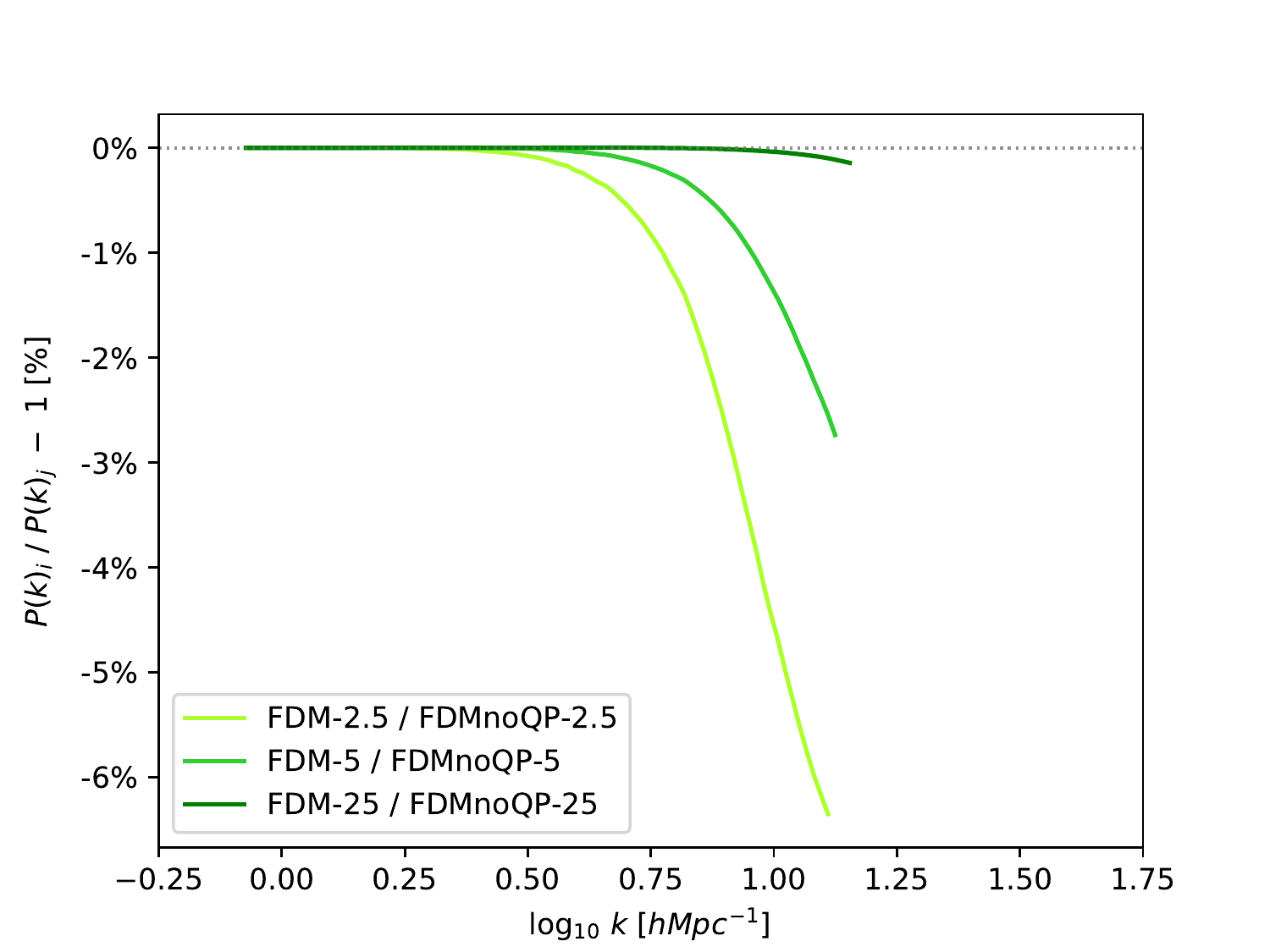}&
\includegraphics[width=0.48\textwidth,trim={0.1cm 0.1cm 1.1cm 1.3cm},clip]{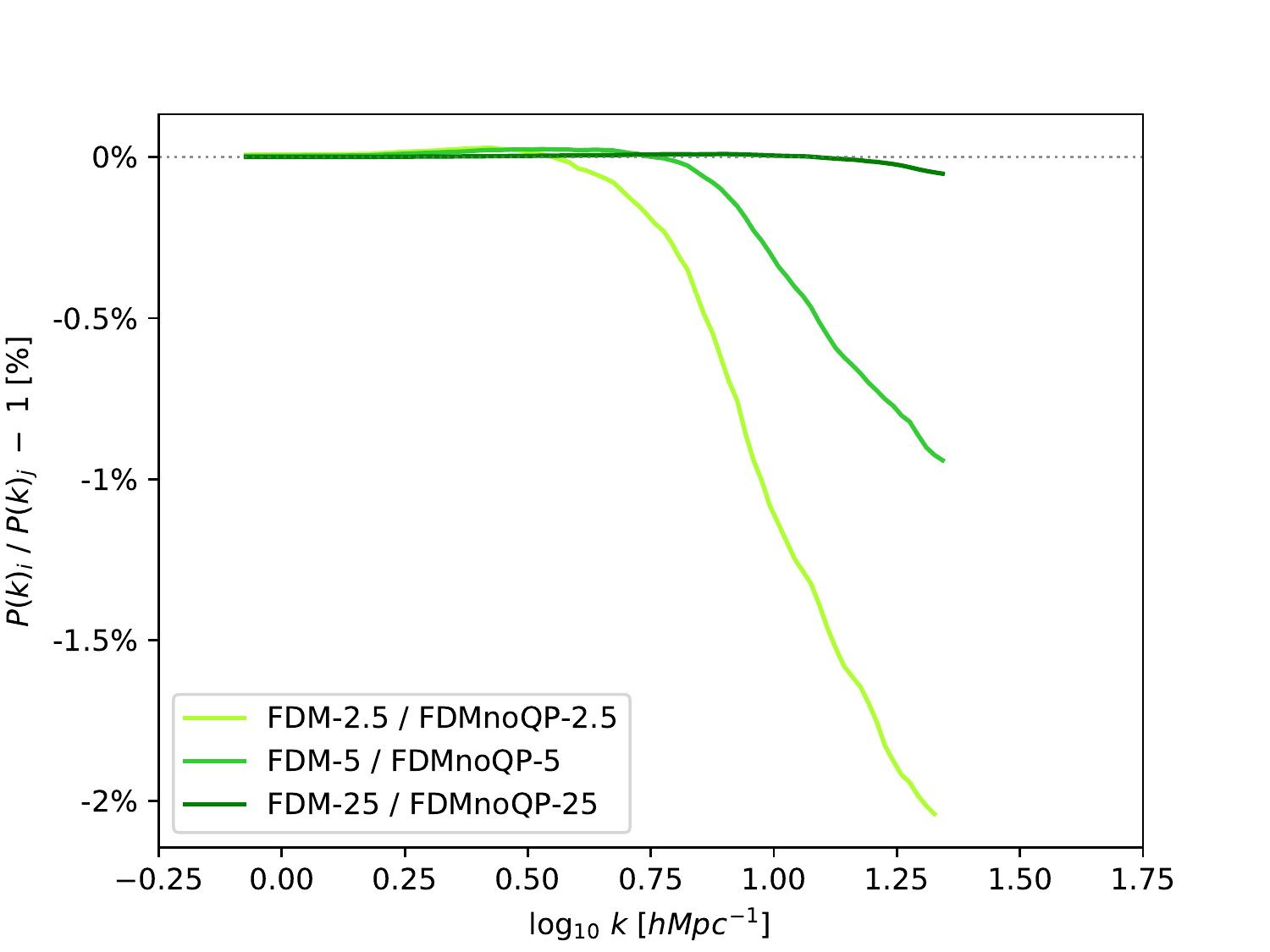}\\
\\
$z=3.6$ & $z=1.8$ \\
\includegraphics[width=0.48\textwidth,trim={0.1cm 0.1cm 1.1cm 1.3cm},clip]{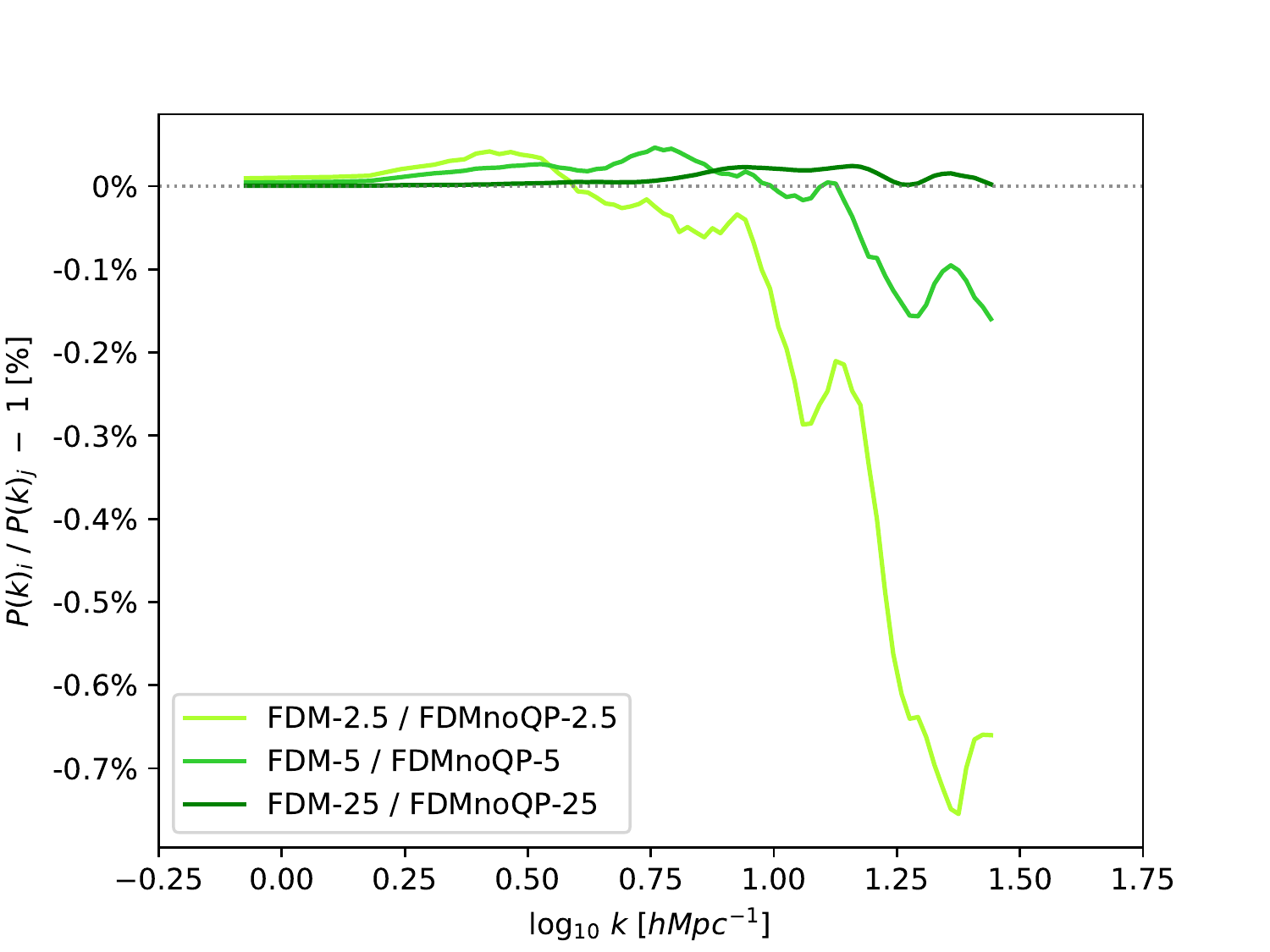}&
\includegraphics[width=0.48\textwidth,trim={0.1cm 0.1cm 1.1cm 1.3cm},clip]{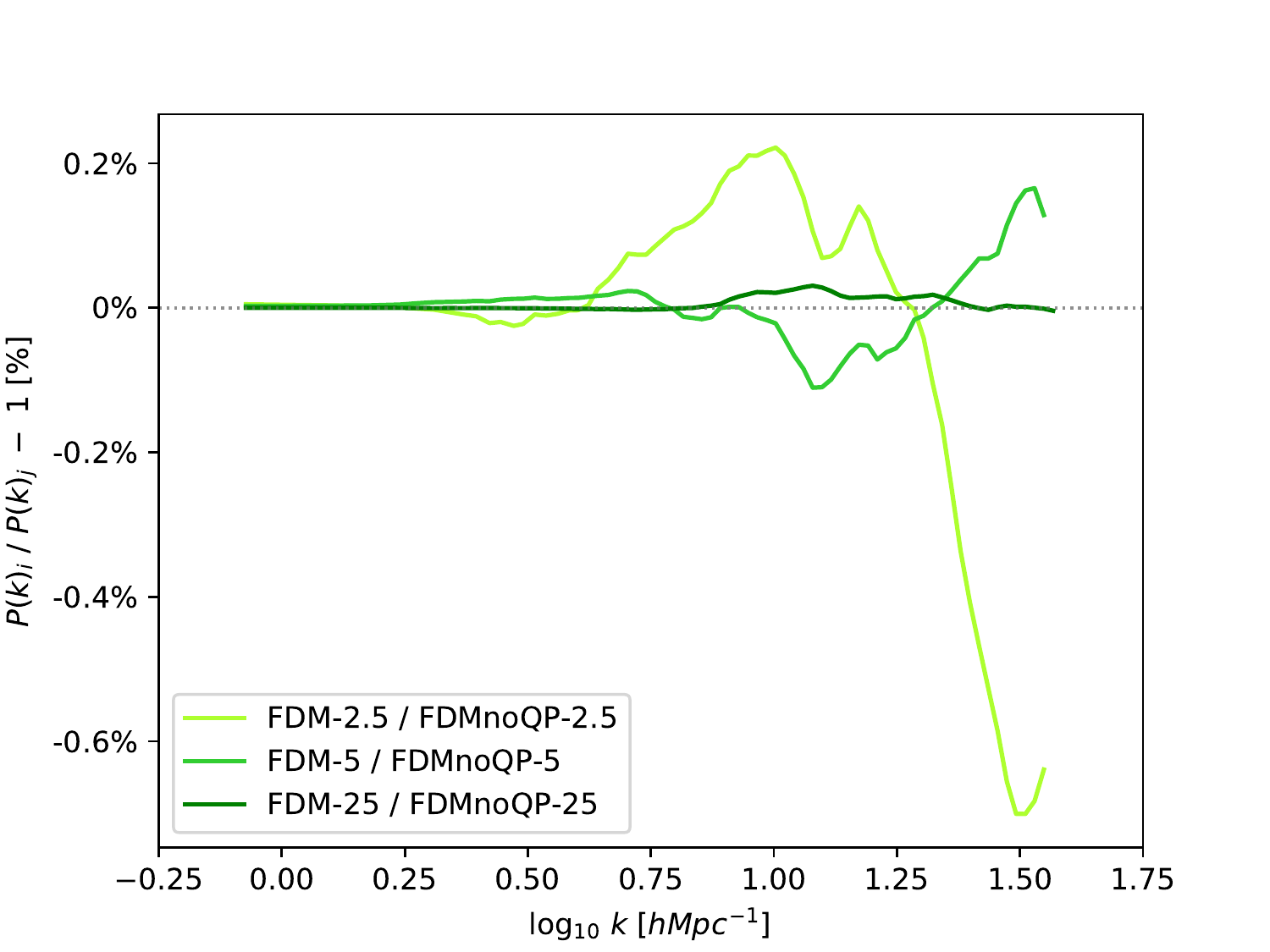}\\
\end{tabular}
\caption{Matter power spectrum percentage differences between FDM simulation and their FDMnoQP counterpart at different redshifts. The difference in power spectrum suppression of having the QP in the dynamics result in a multiplication of $\sim 115\%$ factor of the suppression with respect to LCDM of Fig.~\ref{fig:PS_FDMvsLCDM} at scales $k\sim10\ h Mpc^{-1}$.}
\label{fig:PS_FDMvsFDMnoQP}
\end{figure*}

As already found in the literature \citep[see e.g.][]{Marsh16nl,Nori18}, the evolution of the matter power spectrum shows that the initial suppression --~encoded in the transfer functions used to build up the initial conditions~-- is restored at intermediate scales to the unsuppressed level, eventually, by the non-linear gravitational evolution.

At the redshifts and scales that are relevant for \LA forest observations, however, the relative suppression with respect to $\Lambda$CDM is still important and ranges from $5-20\%$ for the lowest FDM mass considered.

The relative difference of the matter power spectrum, displayed in Fig.~\ref{fig:PS_FDMvsFDMnoQP}, shows an additional suppression with respect $\Lambda$CDM (by a factor $\approx 1.15$) when the QP is included in the dynamical evolution (i.e. in the comparison between the FDM and the FDMnoQP simulations). This is consistent with the QP full dynamical treatment contributing as an integrated smoothing force that contrasts the gravitational collapse of the otherwise purely collisionless dynamics.

\subsection{Lyman-\texorpdfstring{$\alpha$}{}
forest flux statistics}
\label{sec:la_results}

In order to build our simulated Lyman-$\alpha $ observations we extracted 5000 mock forest spectra from random line-of-sights within the simulated volume. 
The spectra are extracted according to SPH interpolation and the ingredients necessary to build up the transmitted flux are the HI-weighted peculiar velocity, temperature and neutral fraction.
Among the different flux statistics that can be considered, we focus on the flux probability distribution function (PDF) and flux power spectrum. Unless otherwise stated we normalize the extracted flux arrays in order to have the same observed mean flux over the whole sample considered and for all the simulations. In any case, we do find that the scaling factor for the optical depth arrays over the whole simulated volume is 1.6, 1.4 and 1.1 times higher than in the $\Lambda$CDM case in order to achieve the same mean flux for the $m_{22}=2.5$, $5$ and $25$ FDM cases with negligible --~between $1-2\%$~-- differences between the FDMs and FDMnoQP cases.

%In Fig.~\ref{fig:la_pdfs} we show the percentage difference between the simulations that include the QP and those that do not include it -- namely FDM and FDMnoQP -- at $z=5.4$, one of the highest redshift bins in which \LA data are available. The flux and gas probability distribution function ratios are presented, in the top and bottom panel, respectively. The differences are quite small and there is a tendency for having a $2-6 \%$ peak at flux $\sim 0.6-0.8$, i.e. in regions of low transmissivity, that are expected to trace voids.

%The fact that simulations which include the QP display a more peaked PDF compared to the other simulations for this range of fluxes is witnessing the fact that, on average, it is more likely to sample such void environments for the FDM simulations rather than the FDMnoQP ones. Basically, the different PDFs should reflect underlying different gas PDFs at the same redshifts and along the same lines-of-sight.

\begin{figure}
\includegraphics[width=\columnwidth]{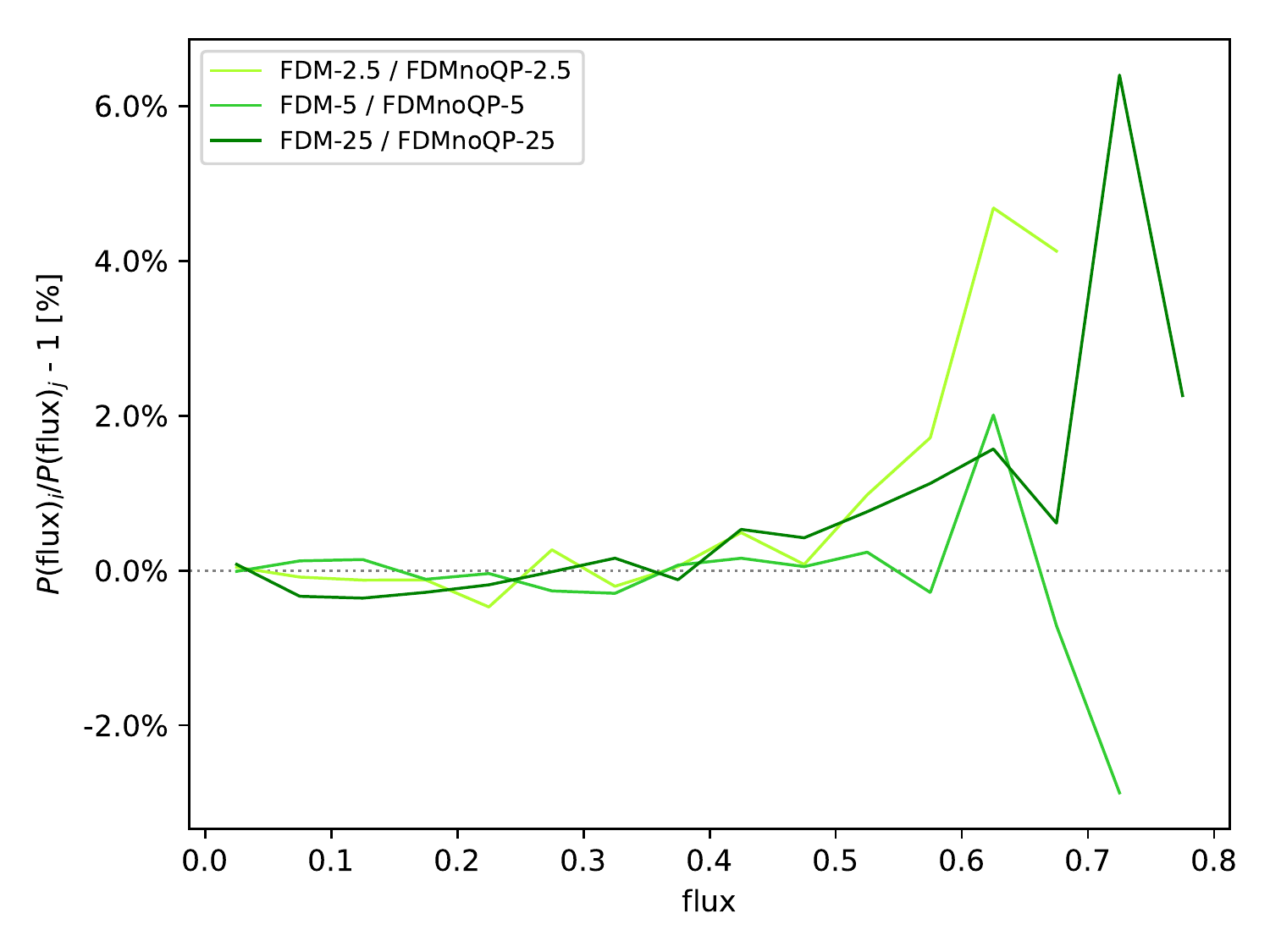}
\includegraphics[width=\columnwidth]{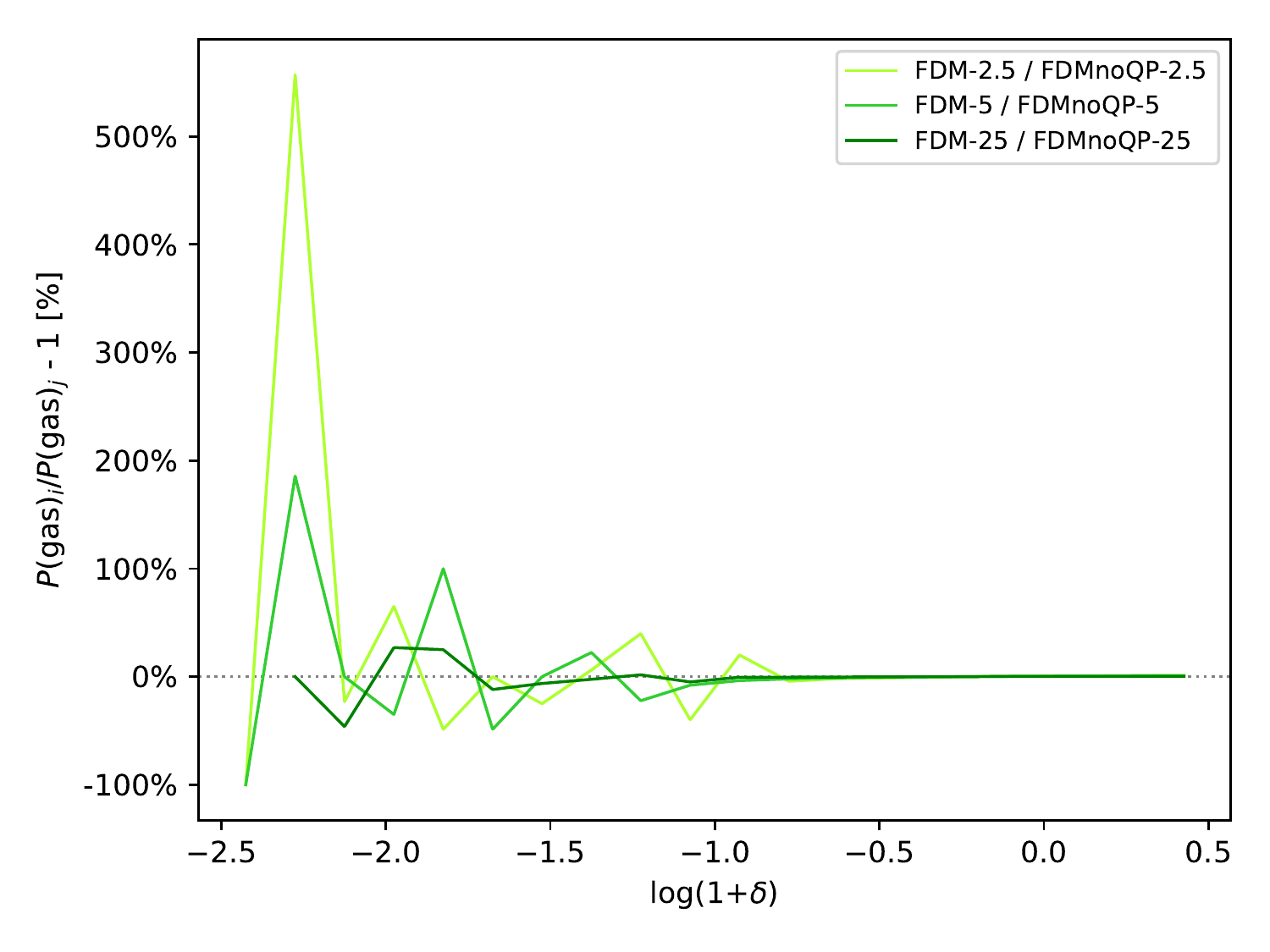}
\caption{Relative differences of the flux PDF (\textit{top panel}) and gas PDF (\textit{bottom panel}) for FDM models with respect to their corresponding FDMnoQP counterparts, at redshift $z=5.4$.}\label{fig:la_pdfs}
\end{figure}

In Fig.~\ref{fig:la_pdfs} we show the flux (\textit{top panel}) and gas (\textit{bottom panel}) PDF ratios between the simulations that include the QP and those that do not include it --~FDM and FDMnoQP, respectively-- at $z=5.4$, one of the highest redshift bins in which \LA data are available.

It is possible to see that there is a $2-6 \%$ peak at flux $\sim 0.6-0.8$, i.e. in regions of low transmissivity that are expected to trace voids. The fact that FDM simulations display a more peaked PDF compared to FDMnoQP ones for this range of fluxes means that, on average, in those models it is more likely to sample such void environments. In fact, the different PDFs should reflect the underlying different gas PDFs at the same redshifts and along the same lines-of-sight. In the bottom panel of Fig.~\ref{fig:la_pdfs}, showing the corresponding gas PDF, it is indeed apparent that in models with FDMs the gas PDF is more skewed towards less dense regions, that are typically associated to high transmission. The effect due to the QP is thus to increase the volume filling factor of regions below the mean density with respect to the corresponding FDMnoQP case.

\begin{figure*}
\begin{tabular}{cc}
\multicolumn{2}{c}{$z=5.4$} \\
\\
\includegraphics[width=0.48\textwidth,trim={0.1cm 0.1cm 1.1cm 1.3cm},clip]{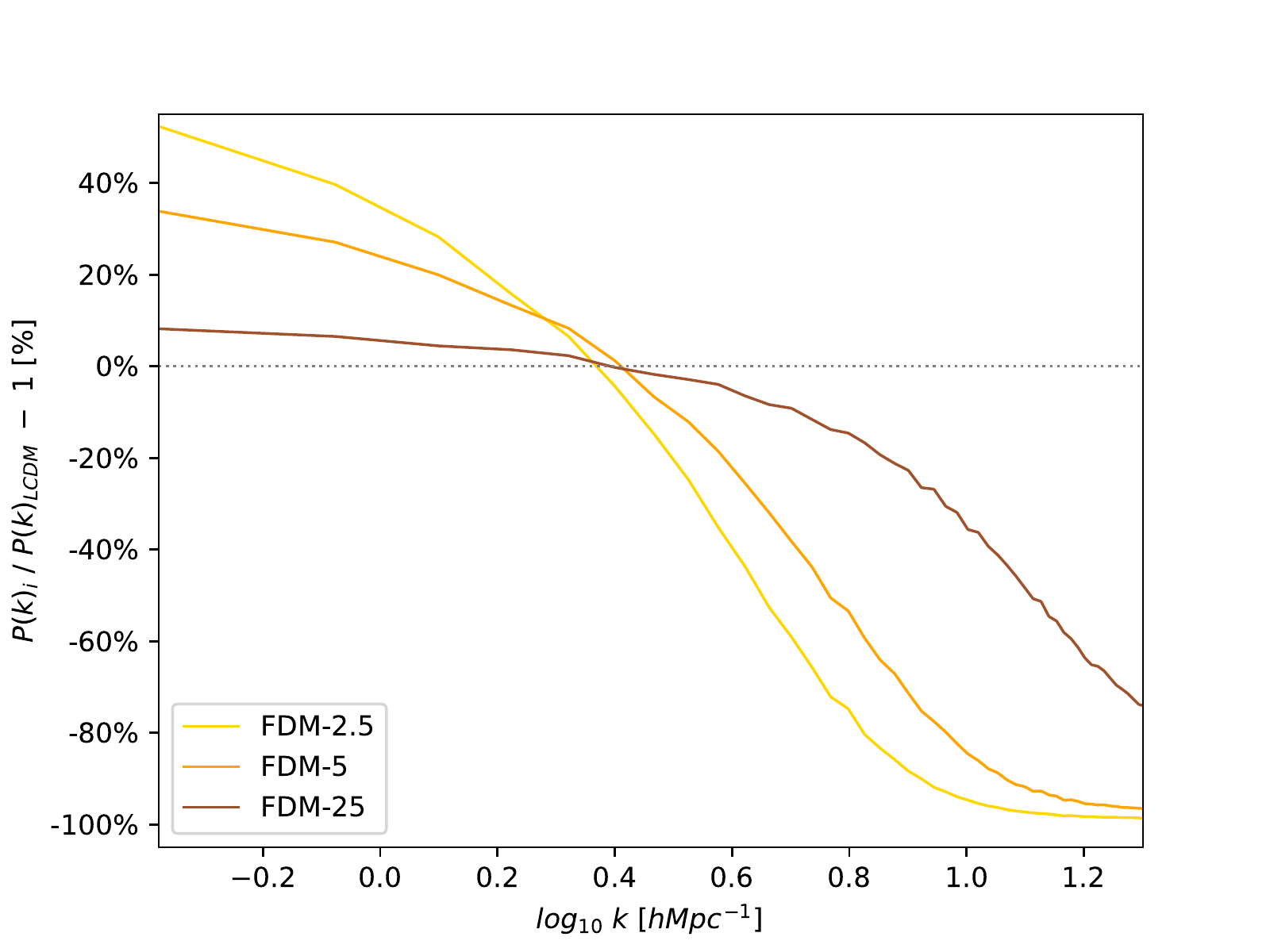}&
\includegraphics[width=0.48\textwidth,trim={0.1cm 0.1cm 1.1cm 1.3cm},clip]{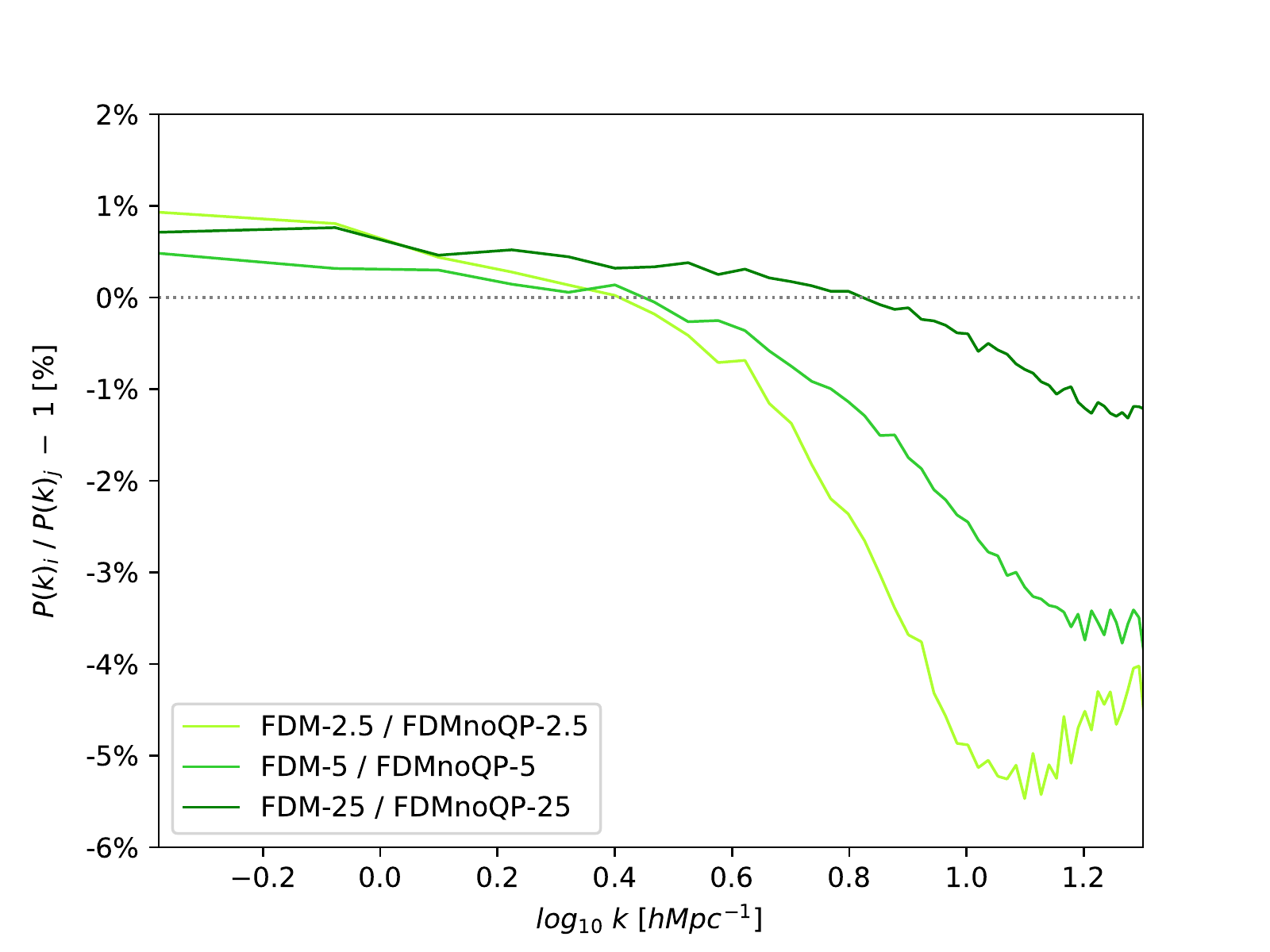}\\
\\
\multicolumn{2}{c}{$z=4.0$} \\
\\
\includegraphics[width=0.48\textwidth,trim={0.1cm 0.1cm 1.1cm 1.3cm},clip]{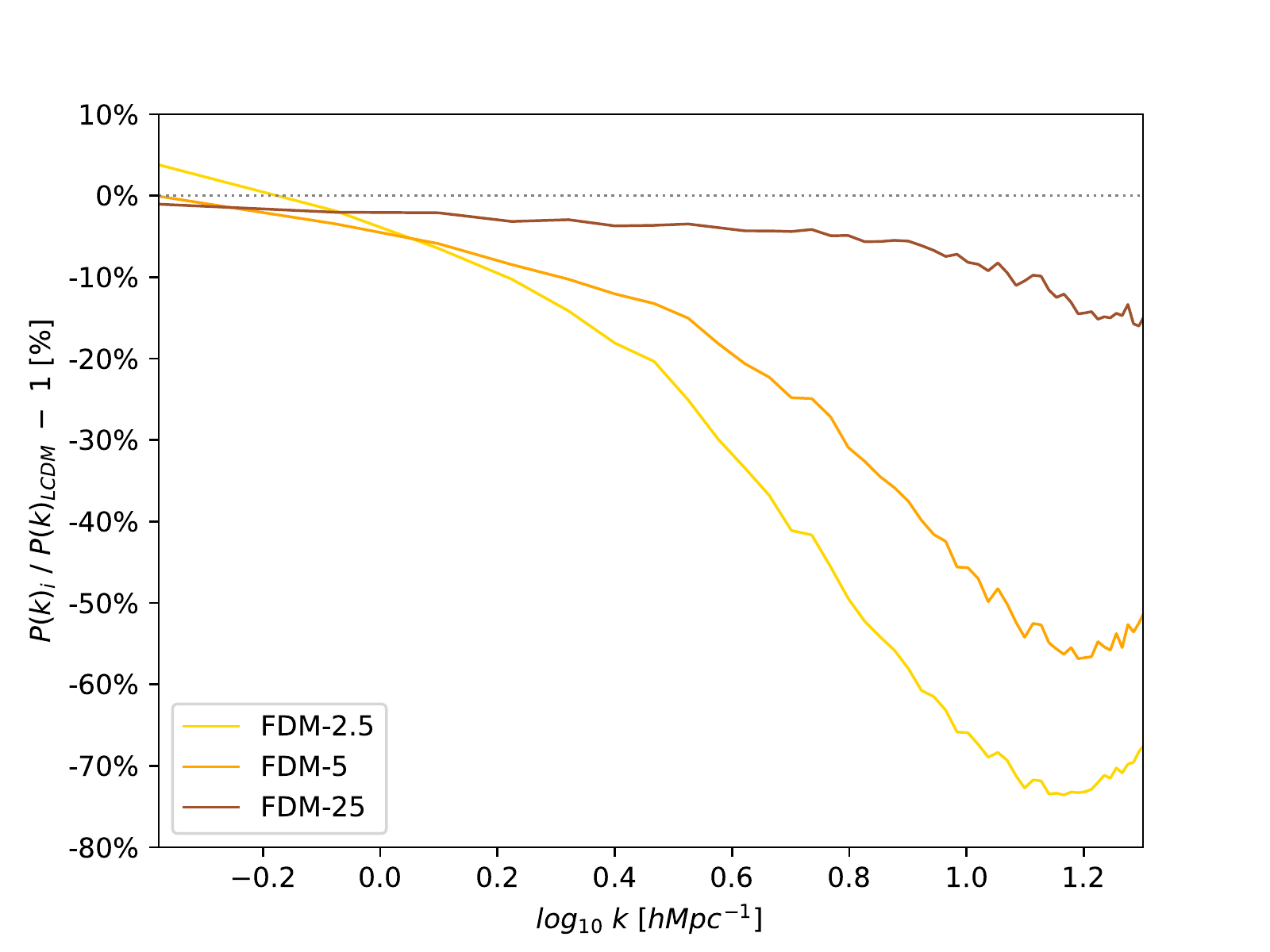}&
\includegraphics[width=0.48\textwidth,trim={0.1cm 0.1cm 1.1cm 1.3cm},clip]{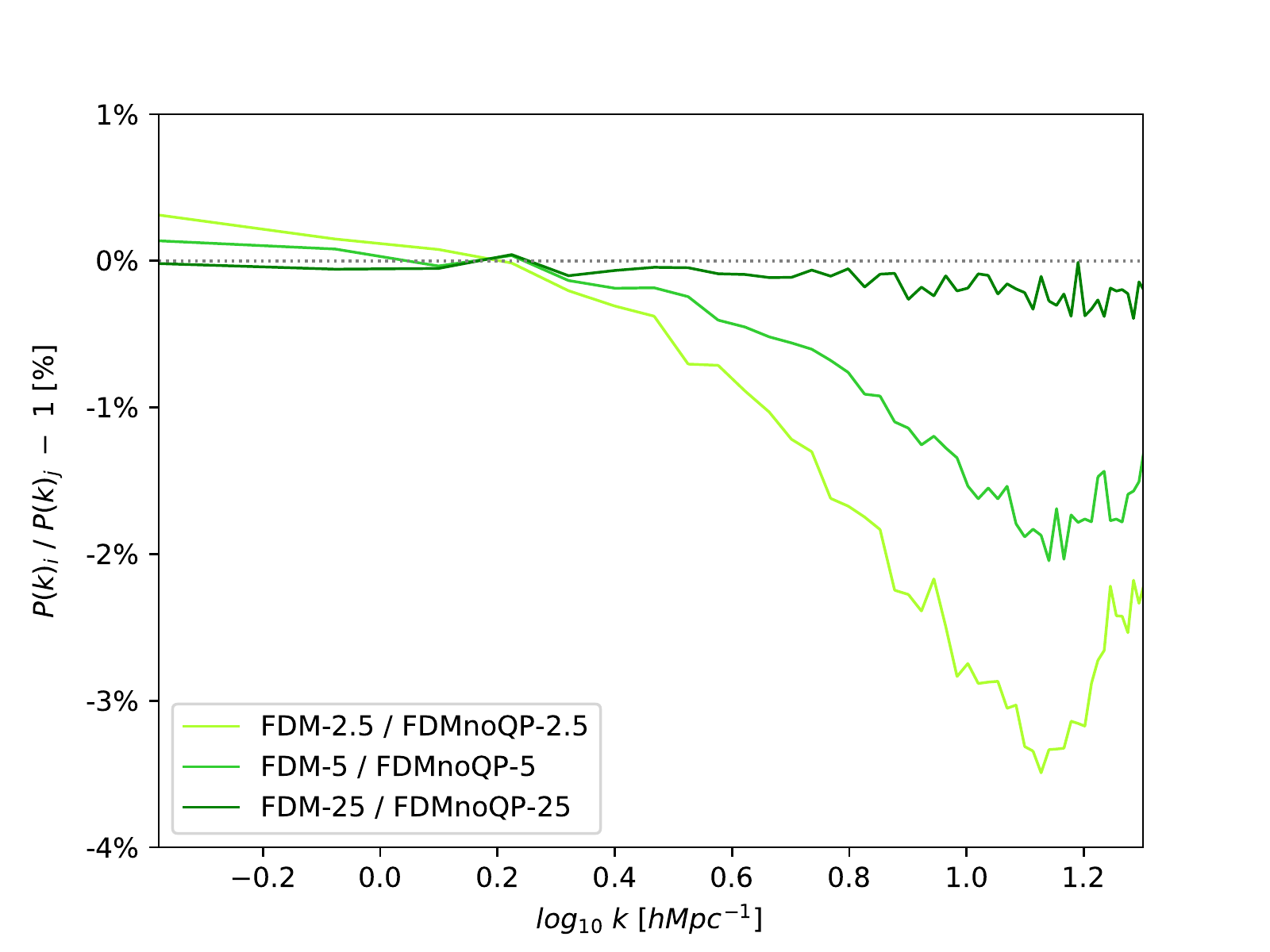}\\
\\
\multicolumn{2}{c}{$z=3.0$} \\
\\
\includegraphics[width=0.48\textwidth,trim={0.1cm 0.1cm 1.1cm 1.3cm},clip]{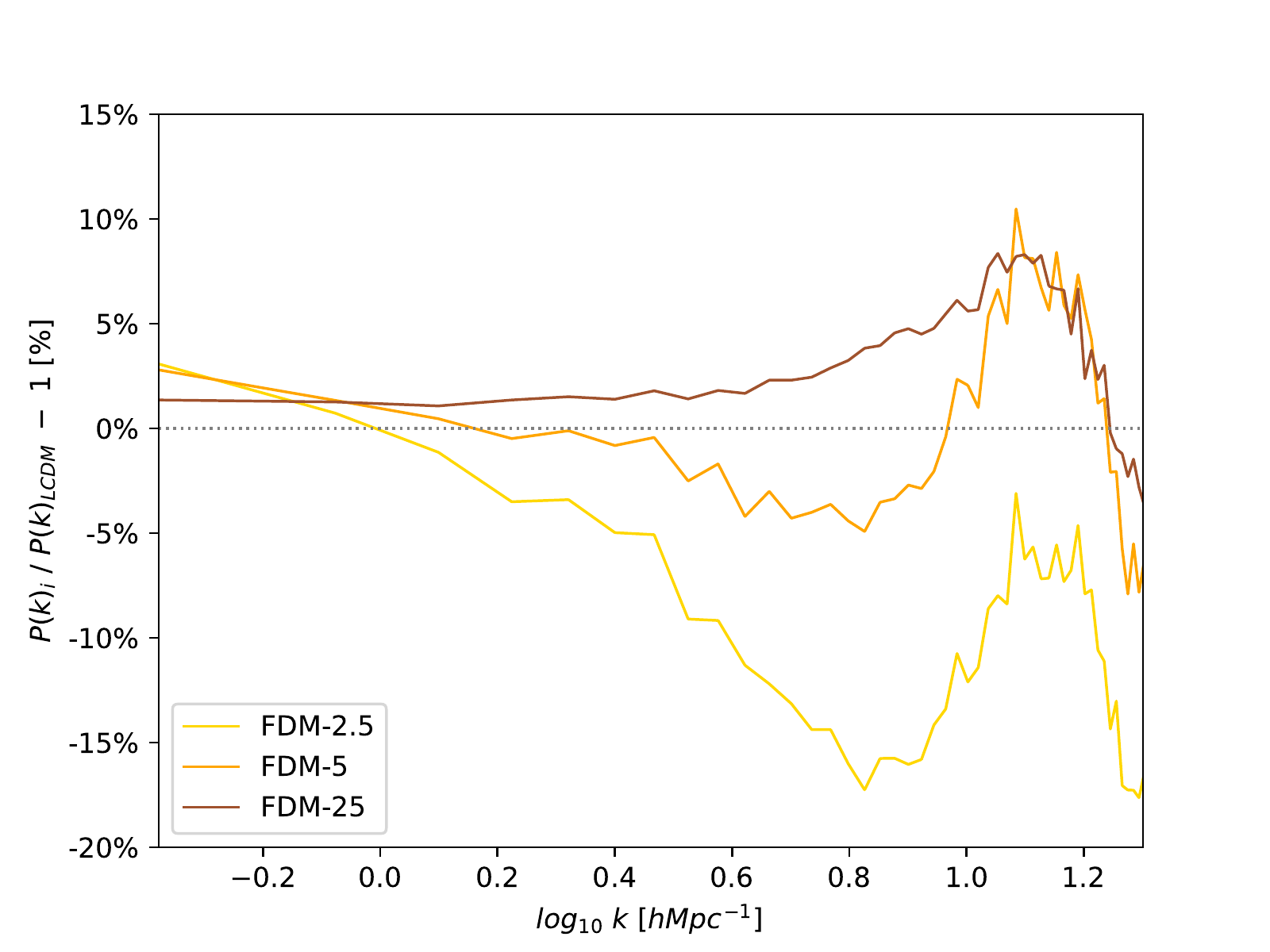}&
\includegraphics[width=0.48\textwidth,trim={0.1cm 0.1cm 1.1cm 1.3cm},clip]{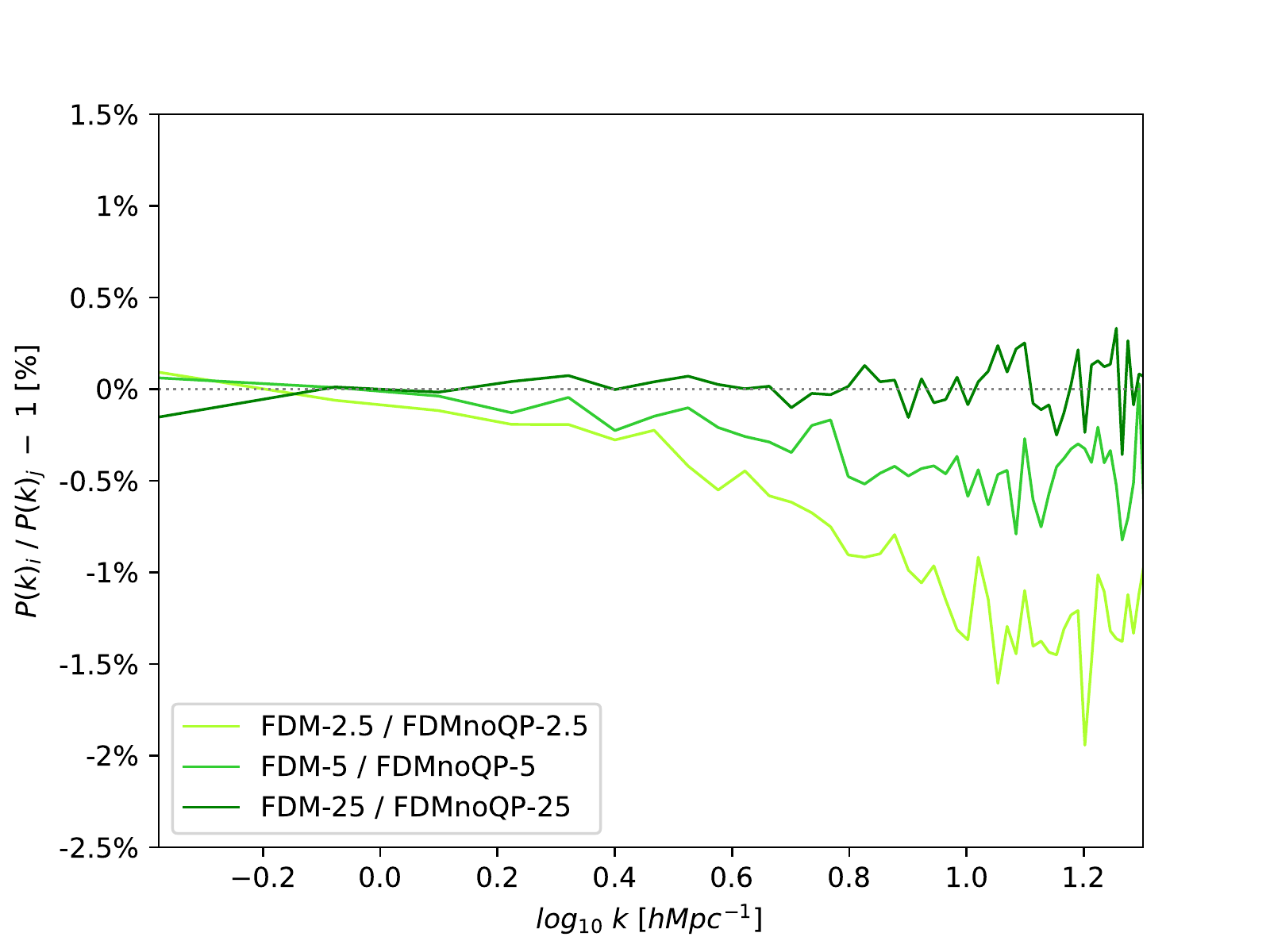}\\
\end{tabular}
\caption{Flux power spectrum comparison between all simulations and LCDM (\textit{left panels}), and between FDM simulation and their FDMnoQP counterparts (\textit{right panels}) at different redshifts.
}
\label{fig:Pflux}
\end{figure*}

\begin{figure}
\centering
\includegraphics[width=0.45\textwidth]{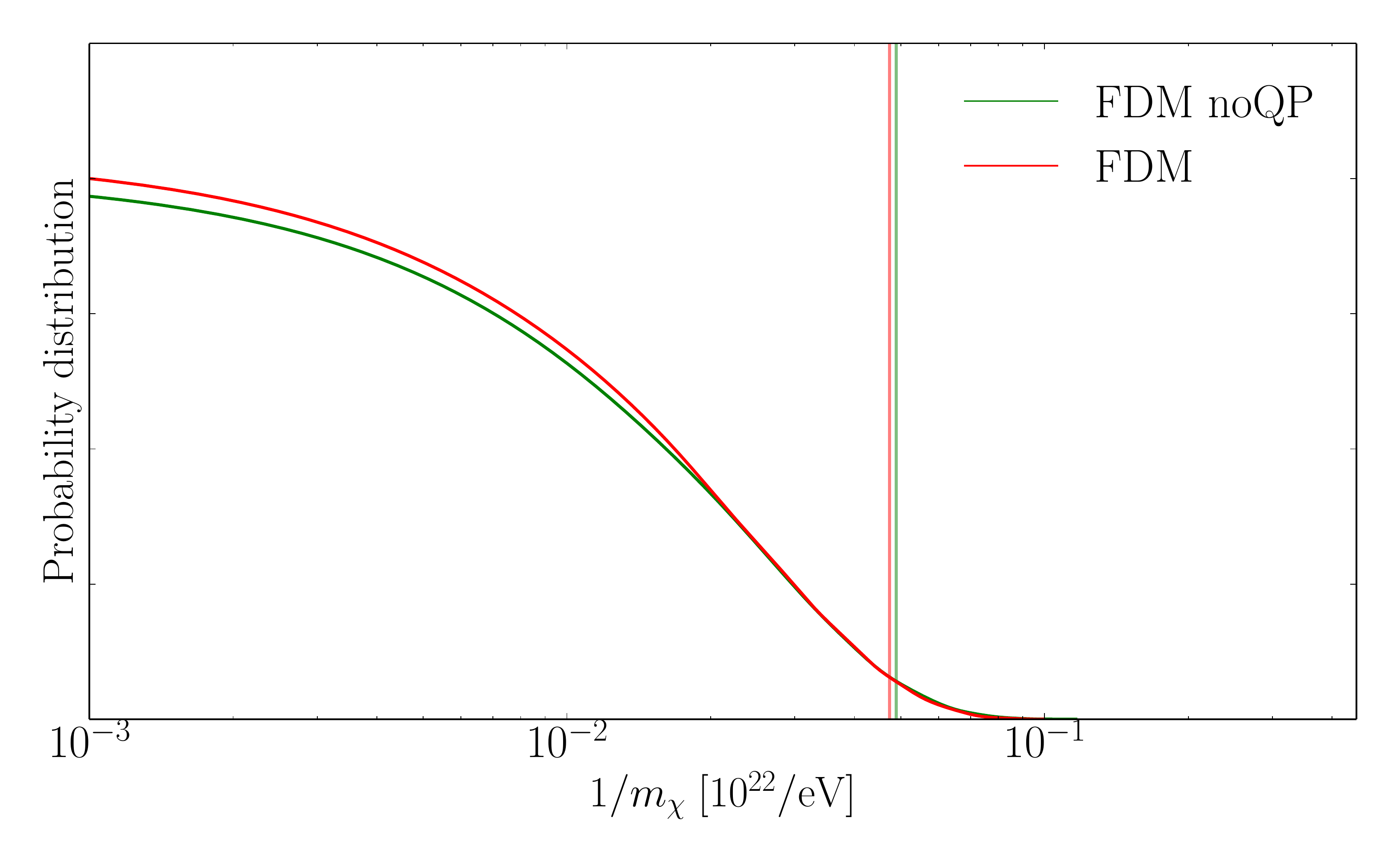}
\caption{Here we plot the marginalised posterior distribution of 1/$m_{\chi}$ from both the analyses performed by \citet{Irsic17} (green lines, without QP) and ours (red lines, with QP). The vertical lines stand for the $2\sigma$ C.L. limits.}
\label{fig:mcmc}
\end{figure}

In Fig.~\ref{fig:Pflux} we plot the percentage difference in terms of flux power spectrum at three different redshifts and for the FDM models, both compared to $\Lambda$CDM (\textit{right panels}) and to the corresponding FDMnoQP case (\textit{left panels}).
The increase of power at $z=5.4$ in the largest scales --~compared to the $\Lambda$CDM case~-- is due to the imposed normalization at the same mean flux, while the evident suppression at small scales is related to the lack of structures at those scales. The comparison with the FDMnoQP set-ups, instead, reveals an additional suppression which is always below the 5\% level for all the masses considered. Since the flux power spectrum is an exponentially suppressed proxy of the underlying density field, these results are consistent with the matter power spectrum results previously shown in Fig.~\ref{fig:PS_FDMvsLCDM} and Fig.~\ref{fig:PS_FDMvsFDMnoQP}.

Since the \LA constraint are calculated by weighting the contribution from all the scales, we expect the bound on $m_{22}$ found in \citet{Irsic17} to change comparably to the additional suppression introduced, that in our case is $~2-3\%$.

This is exactly what can be seen in in Fig.~\ref{fig:mcmc}, where the marginalised posterior distribution of $m_{\chi}$ obtained in the present work is plotted and compared with the results presented in \citet{Irsic17}. The red line refers to our MCMC analysis, whereas the green line corresponds to the results obtained by \citet{Irsic17}. The corresponding vertical lines show the $2\sigma$ bounds on the FDM mass. The $2\sigma$ bound on the FDM mass changes from $20.45 \times 10^{-22}\;\mathrm{eV}$ to $21.08 \times 10^{-22}\;\mathrm{eV}$, which matches with our expectation and confirm that the approximation of neglecting the QP dynamical effects in \citet{Irsic17} was legitimate to investigate the \LA typical scales. The agreement between the sets of results obtained with and without the dynamical QP implementation is evident and is not sensibly affected by varying the assumptions on the IGM thermal history.

This result represents --~to our knowledge~-- the first FDM mass constraint derived from \LA forest observations that accounts for the full non-linear treatment of the QP, which introduces an additional --~albeit not big~-- suppression of the matter power spectrum in the redshift range and comoving scales probed by the \LA forest. The agreement with previous results implies that the non-linear evolution of the large-scale structure and the non-linear mapping between flux and density effectively make up for the additional suppression introduced.

\subsection{Structure characterization}
\label{sec:SC}

\begin{figure*}
\centering
\includegraphics[width=0.48\textwidth,trim={0.1cm 0.1cm 1.5cm 1.3cm},clip]{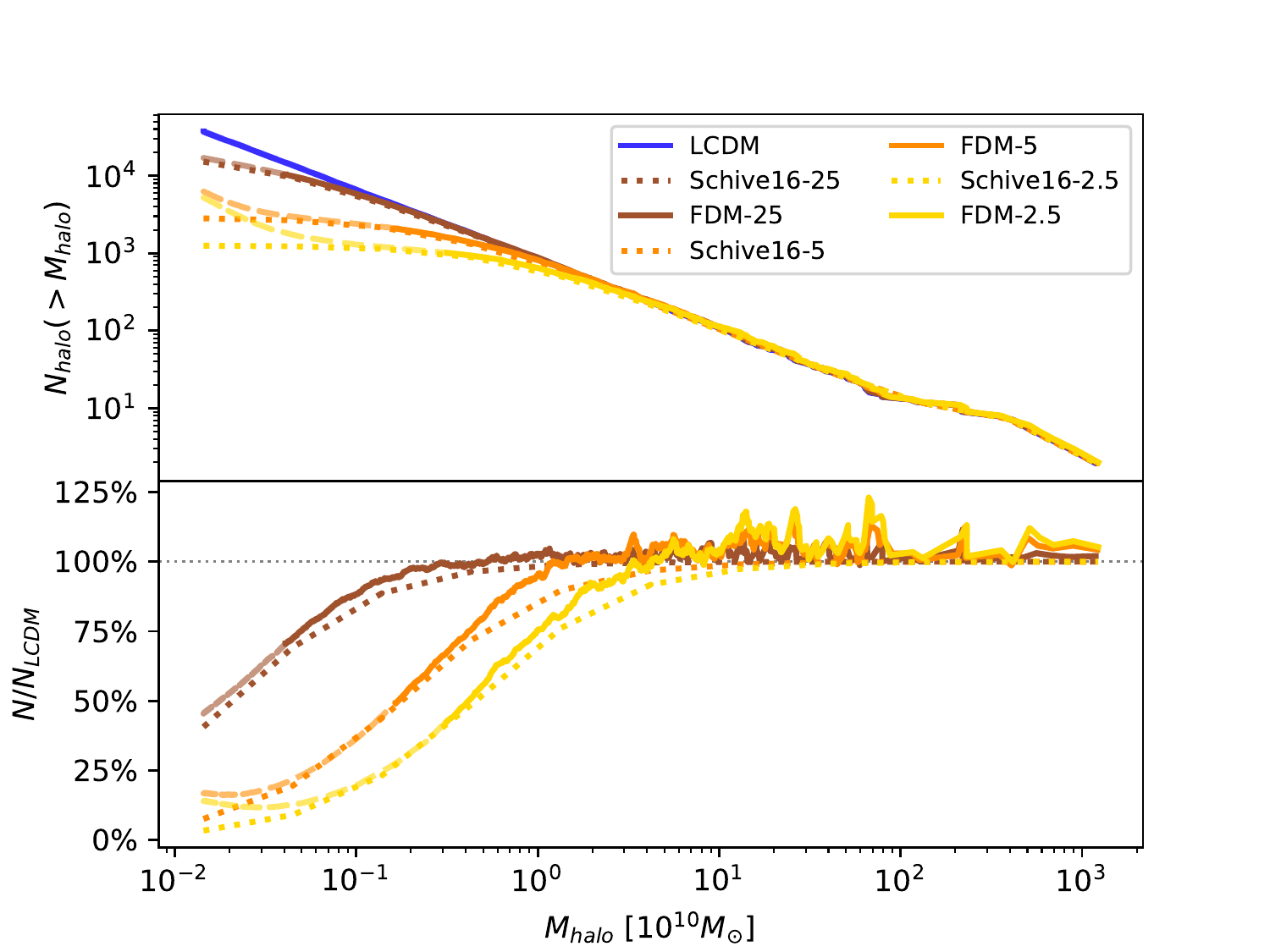}
\includegraphics[width=0.48\textwidth,trim={0.1cm 0.1cm 1.5cm 1.3cm},clip]{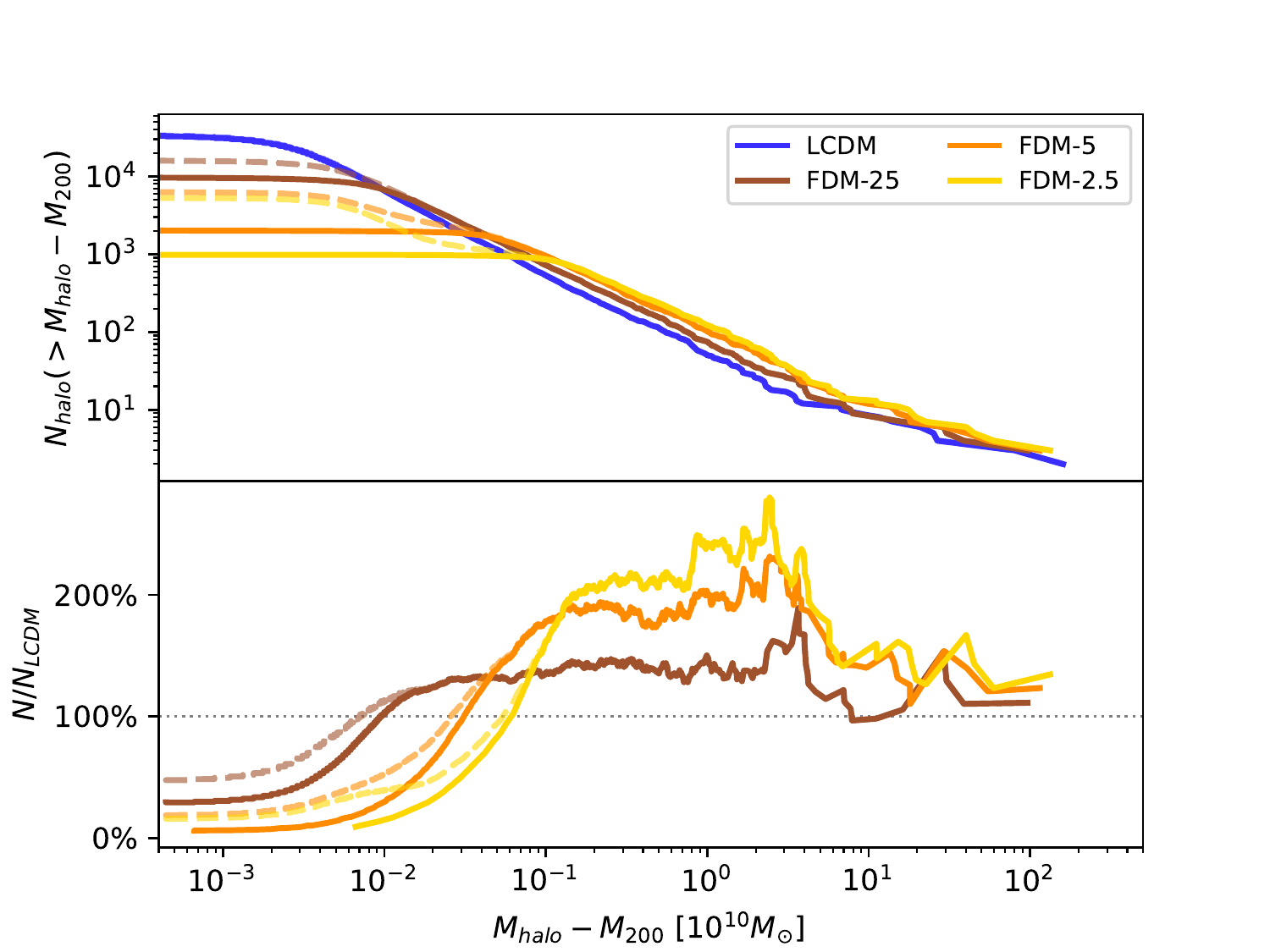}
\includegraphics[width=0.48\textwidth,trim={0.1cm 0.1cm 1.5cm 1.3cm},clip]{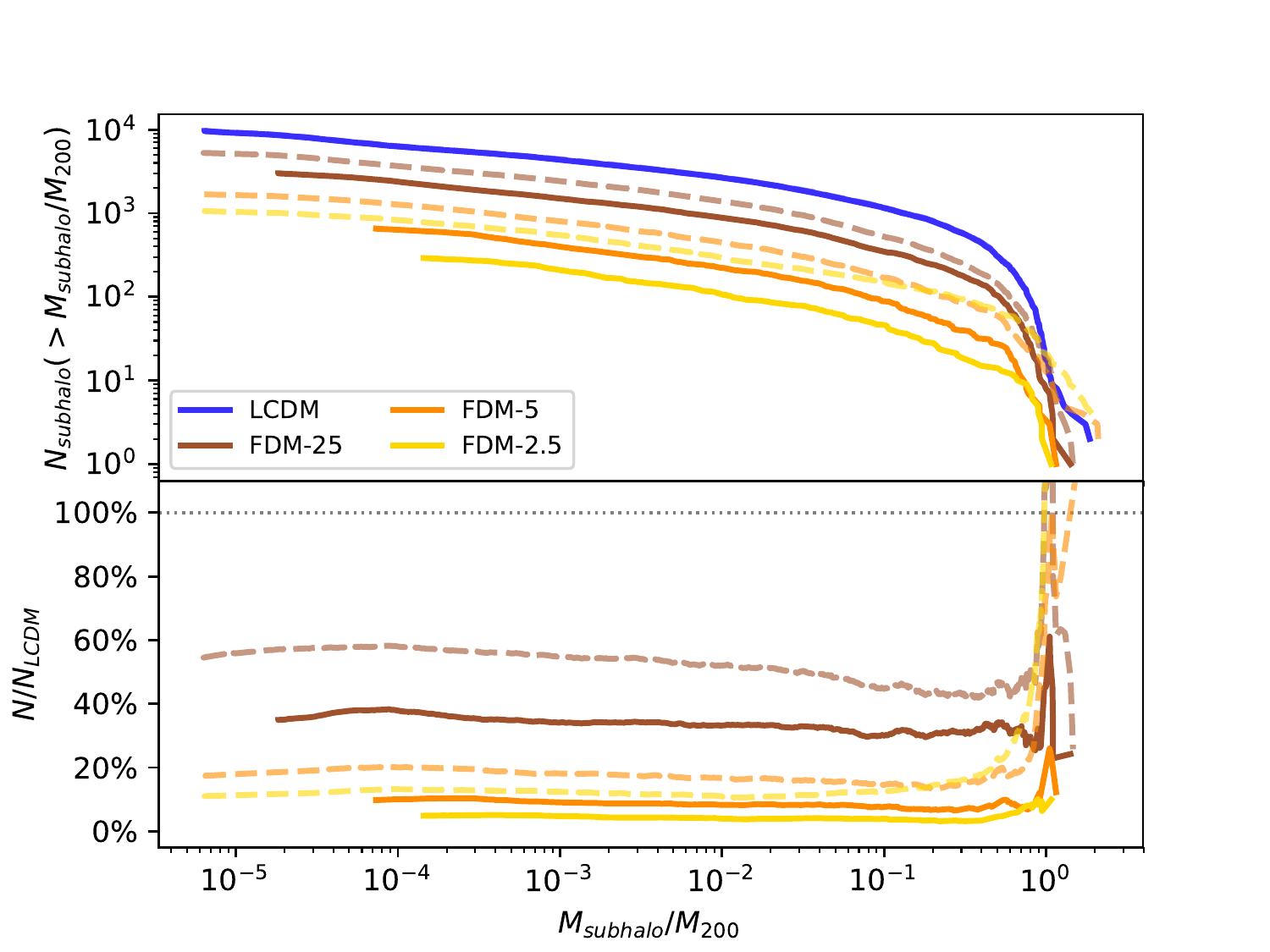}
\includegraphics[width=0.48\textwidth,trim={0.1cm 0.1cm 1.5cm 1.3cm},clip]{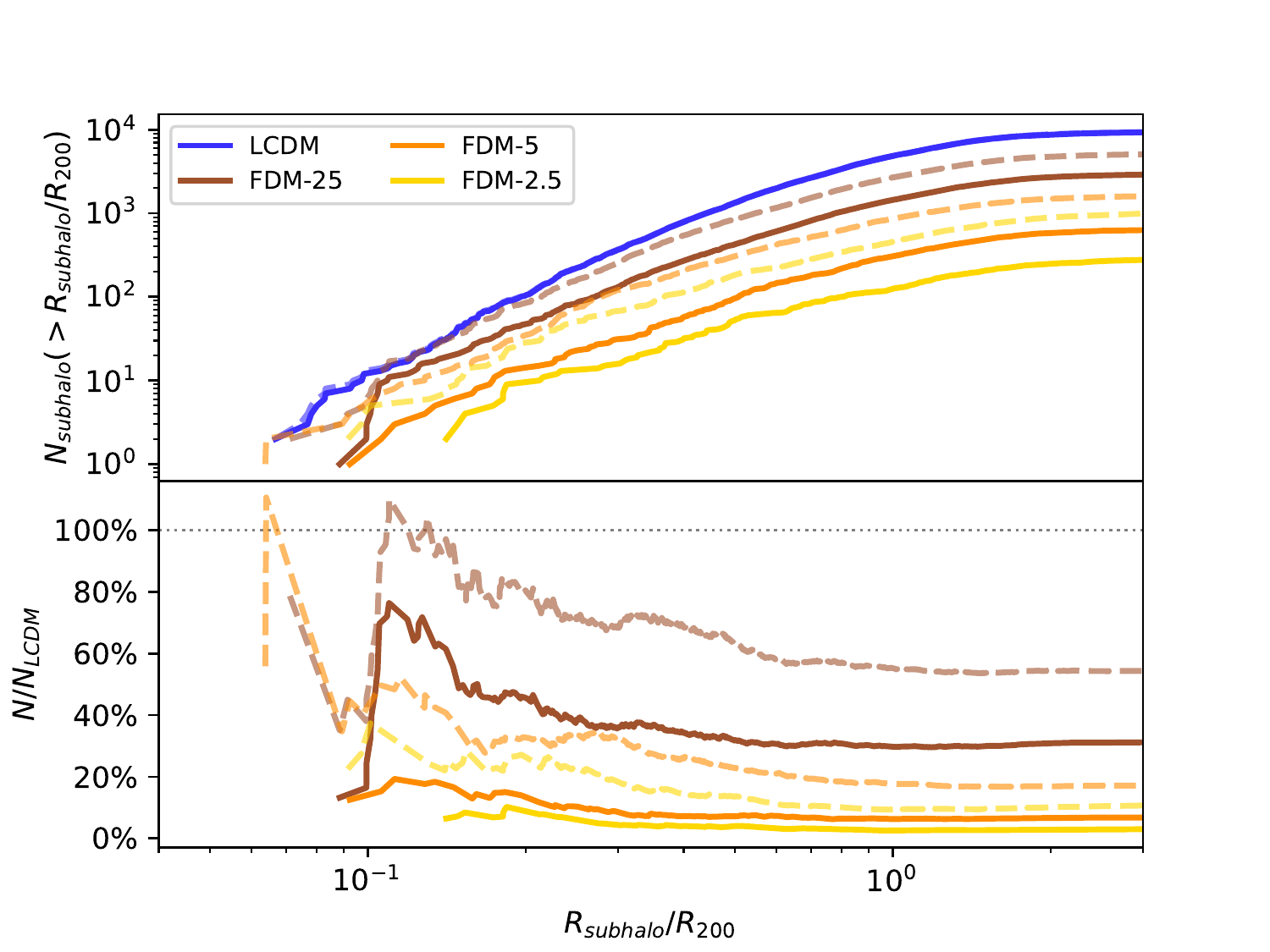}
\caption{Properties of the halo and subhalo samples at $z=0$, with (\textit{dashed lines}) and without (\textit{solid lines}) including the haloes marked as spurious as described in Sec.~\ref{sec:NF}. In particular, the cumulative distributions of halo mass (\textit{top left panel}), the halo mass outside $R_{200}$ (\textit{top right panel}), the subhalo-halo relative mass (\textit{bottom left panel}) and the subhalo-halo distance (\textit{bottom right panel}) are displayed. The fitting functions of the cumulative halo mass distribution of \citep{Schive16} of Eq.~\ref{eq:schivefit} are plotted for reference --~\textit{dotted} line in the \textit{top left panel}~--.}
\label{fig:HALOGROUP}
\end{figure*}

The statistical properties of the genuine haloes belonging to each simulation are summarised in Fig.~\ref{fig:HALOGROUP}, where we display the cumulative halo mass function (top right panel), the halo mass outside $R_{200}$ (top left panel) --~where $R_{200}$ identifies the distance from the halo centre where the density is 200 times the critical density of the Universe and $M_{200}$ the mass contained within a $R_{200}$ radius sphere~--, the subhalo mass function (bottom left panel), and the subhalo radial distribution (bottom right panel). In order to highlight the impact of numerical fragmentation and simplify the comparison of the different models to $\Lambda$CDM, relative ratios are displayed in the bottom panels and shaded lines represent the distribution of the full halo sample, i.e. including also spurious haloes.

The analytical fit used by  \citet{Schive16} to parameterize the cumulative HMF drop of the FDM models with respect to $\Lambda$CDM 
\begin{equation}
\label{eq:schivefit}
N(>M)_{\text{FDM}}= \int_M^{+\infty}\de{M}N_{\text{CDM}}\left[ 1 + \left(\frac{M}{M_0}\right)^{-1.1} \right]^{-2.2} dM
\end{equation}
with $M_0 = 1.6 \times 10^{10} m_{22}^{-4/3}\ M_{\odot}$, are plotted as reference --~one for each FDM mass~-- in the top left panel of Fig.~\ref{fig:HALOGROUP} (dotted lines).

As expected, we find that the number of small mass subhaloes is drastically reduced in the FDM models and the cumulative distributions depart from $\Lambda$CDM at higher and higher masses as the $m_{\chi}$ mass decreases. The values at which the drop occurs are approximately $5\times10^{10} M_{\odot}$, $2.5\times10^{10} M_{\odot}$ and $5\times10^{9} M_{\odot}$ for values of $m_{22}$ of $2.5$, $5$ and $25$, respectively: this suggest a linear trend of the \textit{threshold} mass
\begin{equation}
M_t \simeq 5 \times 10^{10} M_{\odot} \left( \frac {2.5} {m_{22}} \right)
\end{equation}
describing the approximate mass below which the number of haloes starts decreasing with respect to $\Lambda$CDM.

Looking at the distribution of subhaloes masses as compared to their associated primary halo $M_{200}$ and the radial distribution to $R_{200}$, it is evident how the numerous small subhaloes in $\Lambda$CDM, far from the gravitational centre of the main halo, are the ones that were not able to form in a FDM universe.

The haloes that have masses above $M_t$ not only have been able to survive the disrupting QP action up to redshift $z=0$, but the cumulative distribution shows how they also gained extra mass, at the smallest (sub)halo expenses. This is confirmed by the cumulative distribution of the primary structures $N(>Mtot-M_{200})$, representing the mass accumulated outside the $R_{200}$ radius, which is systematically higher with respect to $\Lambda$CDM case as the FDM mass lowers --~up to peaks of $200\%$ ratio for the lowest $m_{22}$~--: this is consistent with the picture of bigger primary haloes accreting the mass of un-collapsed smaller subhaloes that did not form.

The fitting function of Eq.~\ref{eq:schivefit} is consistent with the scale of the drop of the HFM, which is indeed expected to be almost redshift independent, since it is predominantly given by the initial PS cut-off \citep{Hu00}. However, it fails to reproduce the data on two levels: on one hand it does not recover the slope of the cumulative distribution --~especially in the mass range close to $M_t$ where the HMF departs from $\Lambda$CDM~-- and, on the other hand, does not account for the mass transfer from smaller haloes, unable to collapse due to QP repulsive interaction, to bigger ones, that accrete the more abundant available matter from their surroundings. The discrepancies between the \citet{Schive16} fitting function and our results are probably due to the fact that the former is based on simulations with approximated FDM dynamics and evolved only to redshifts $z=4$, thus representing a different collection of haloes that are, moreover, in an earlier stage of evolution.% that have all masses $M\lesssim3 \times 10^{11}$. %The lack of the QP dynamical treatment -- in the comparison between our FDMs and FDMnoQPs HMF -- does not play a key role in this discrepancy, since it does not majorly affect the most massive end at $z=0$, but rather there is an overestimation of haloes in around the mass where the cumulative HMF drops, where for example the difference for the FDM-2.5 vs FDMnoQP-2.5 is of 6\%.

Therefore, the analysis of the aggregated data of cumulative distributions of genuine haloes in each simulation lead us to conclude that formation, the evolution and the properties of a FDM halo subject to the real effect of the QP --~as compared to the FDMnoQP approximation~-- can follow three general paths depending on its own mass and on the mass of the FDM boson: if the halo mass is $M \ll M_t$, there is high chance that the halo does not form at all since gravitational collapse is prevented by the QP; if $M\gtrsim M_t$, the halo can be massive enough to form but its properties will be affected by the QP --~especially on its internal structure, as we will see below~--, while for $M \gg M_t$ the halo is not severely affected by the QP, and will simply accrete more easily un-collapsed mass available in its surroundings.

\bigskip

In order to study in more detail the impact of FDM on the halo properties and structures, we divided our common sample, that by construction collects the haloes across all the simulations that share the same $\Lambda$CDM match (as described in detail in Section~\ref{sec:ISHM}), in three contiguous mass ranges. Let us remind that matching haloes have similar but not necessarily equal mass, so mass intervals are to be referred to the $\Lambda$CDM halo mass; the other matching haloes belonging to the FDM simulations are free to have lower and higher mass, compatibly with the limit imposed by the $\tilde{M}$ parameter of the common sample selection procedure. The common sample low mass end is clearly limited by the FDM-2.5 model, since it is the one with higher $M_t$, below which haloes have statistically lower chance to form. The three mass ranges are $[0.5-4], [4-100], [100-4000] \times 10^{10} M{\odot}$, in order to be compatible with the three halo categories described in the previous paragraph for the FDM-2.5 model, being $M_t(m_{22}=2.5) \sim 5 \times 10^{10} M{\odot}$

For all the matching haloes considered, we have tested the sphericity distribution, the halo volume and the total halo mass with respect to $\Lambda$CDM, as well as the radial density profiles.

\begin{figure*}
\begin{tabular}{ccc}
$[0.5-4]\times 10^{10} M{\odot}$ & $[4-100]\times 10^{10} M{\odot}$ & $[100-4000] \times 10^{10} M{\odot}$ \\
\\
\includegraphics[width=0.32\textwidth,trim={0.1cm 0.8cm 1.4cm 1cm},clip]{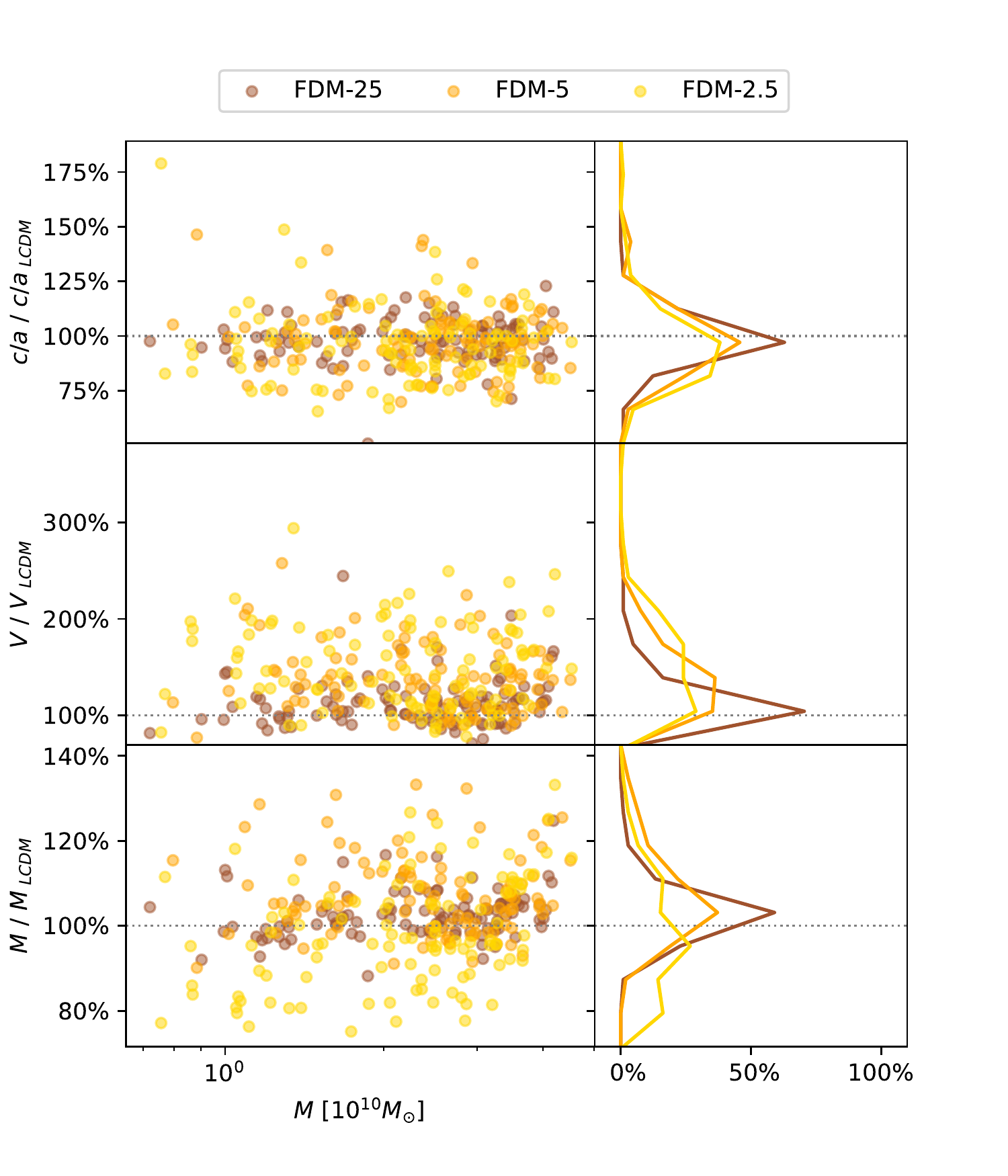} &
\includegraphics[width=0.32\textwidth,trim={0.1cm 0.8cm 1.4cm 1cm},clip]{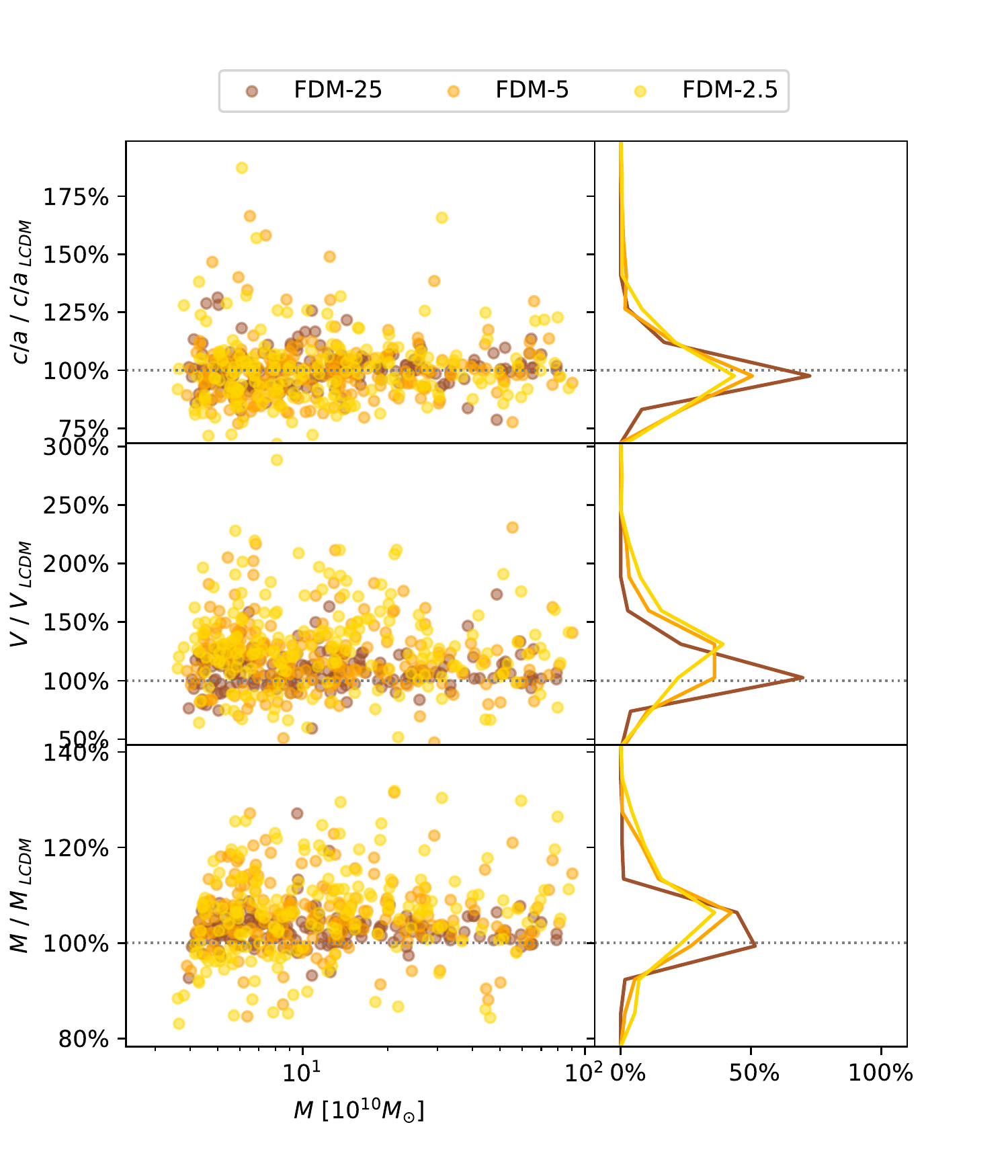} &
\includegraphics[width=0.32\textwidth,trim={0.1cm 0.8cm 1.4cm 1cm},clip]{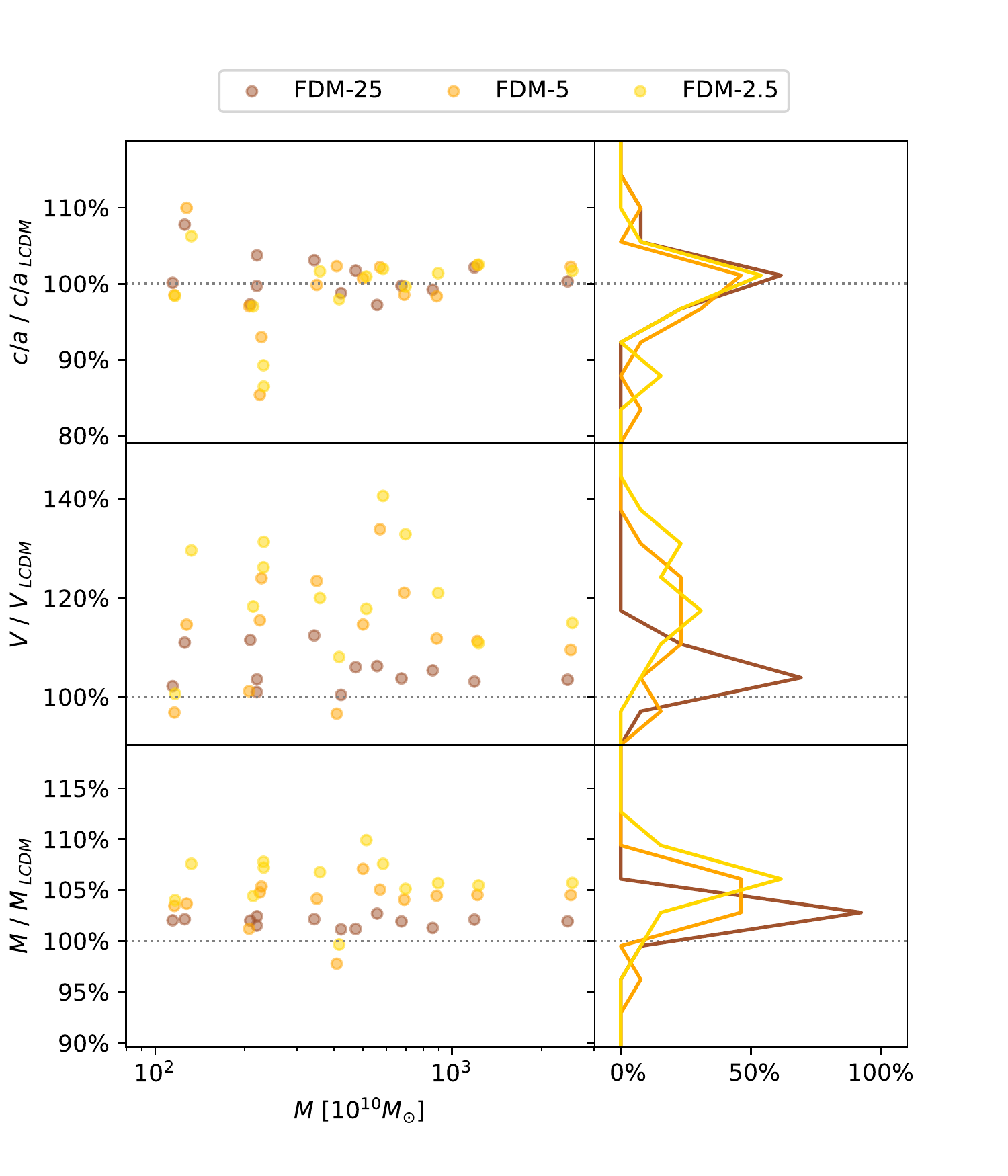} \\
\includegraphics[width=0.32\textwidth,trim={0.1cm 0.3cm 1.4cm 1.5cm},clip]{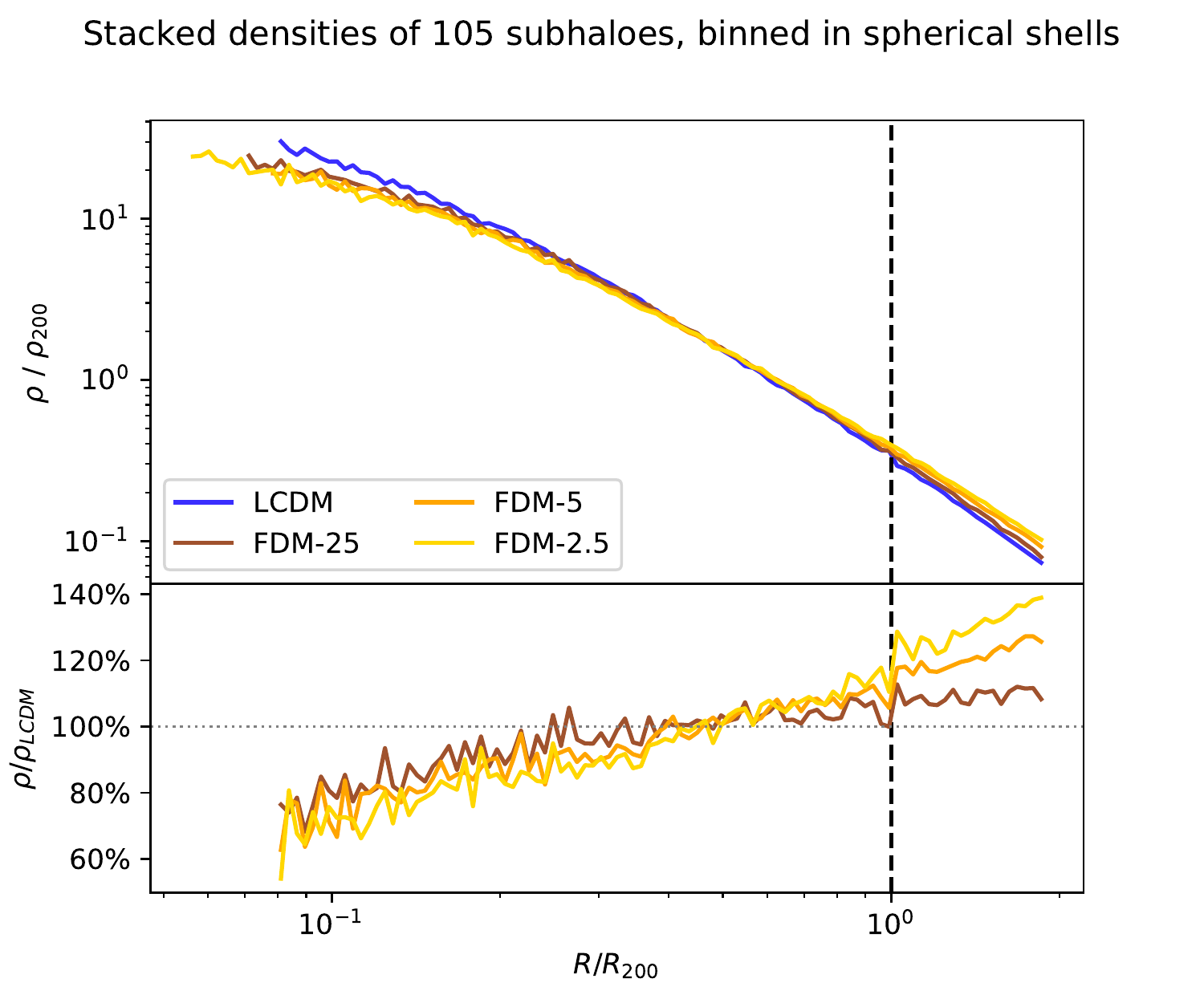} &
\includegraphics[width=0.32\textwidth,trim={0.1cm 0.3cm 1.4cm 1.5cm},clip]{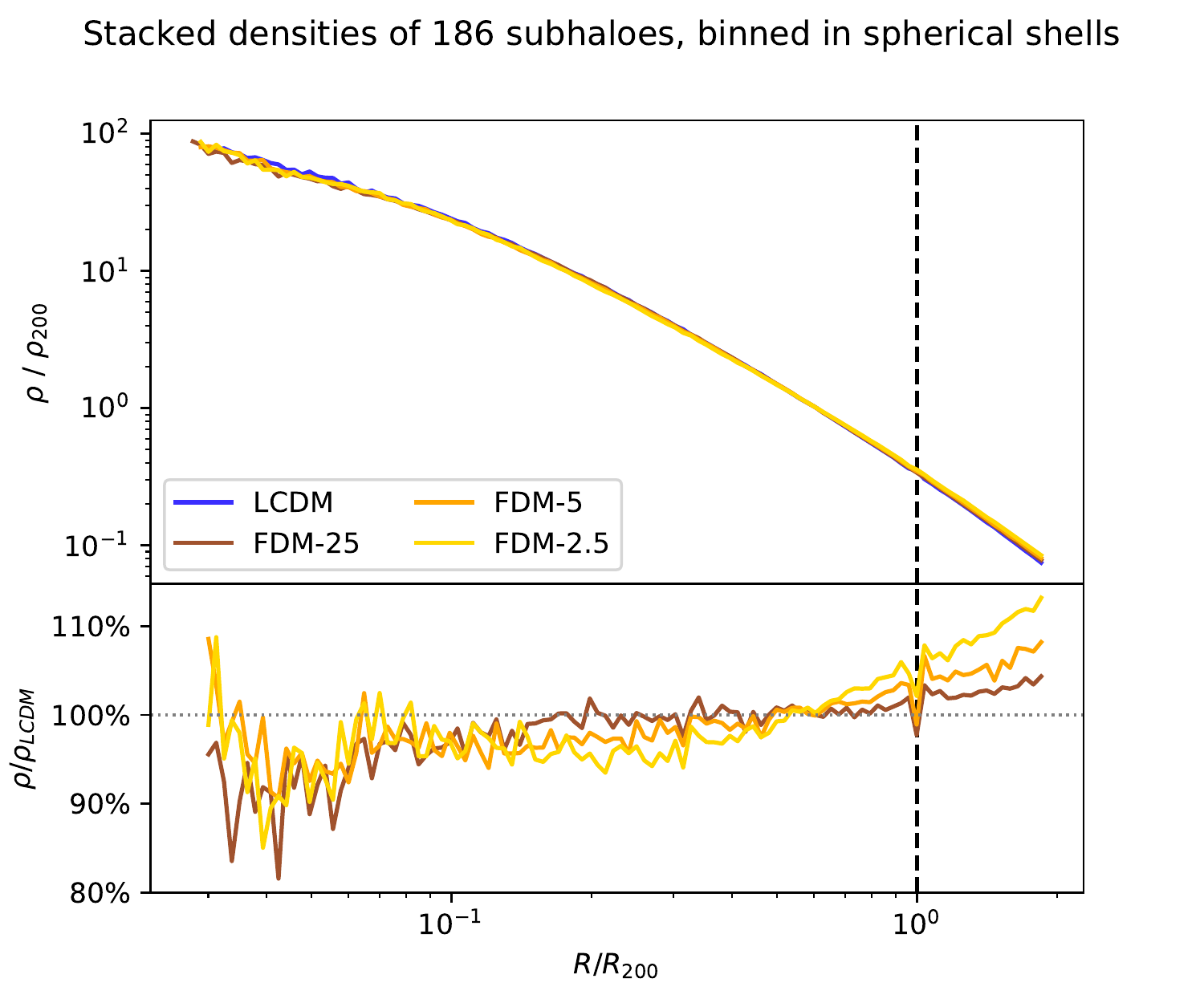} &
\includegraphics[width=0.32\textwidth,trim={0.1cm 0.3cm 1.4cm 1.5cm},clip]{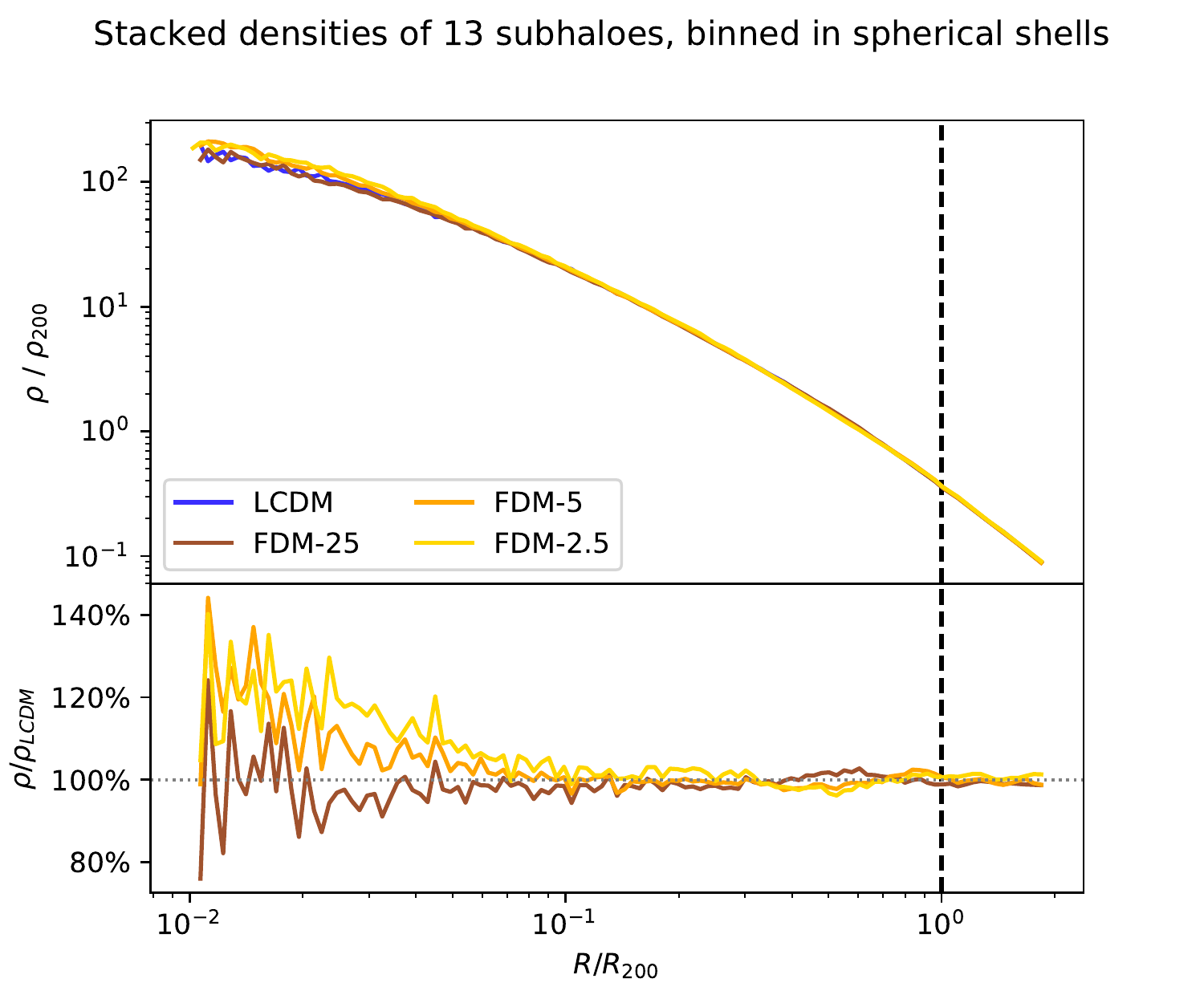} \\
\includegraphics[width=0.32\textwidth,trim={0.1cm 0.3cm 1.4cm 1.5cm},clip]{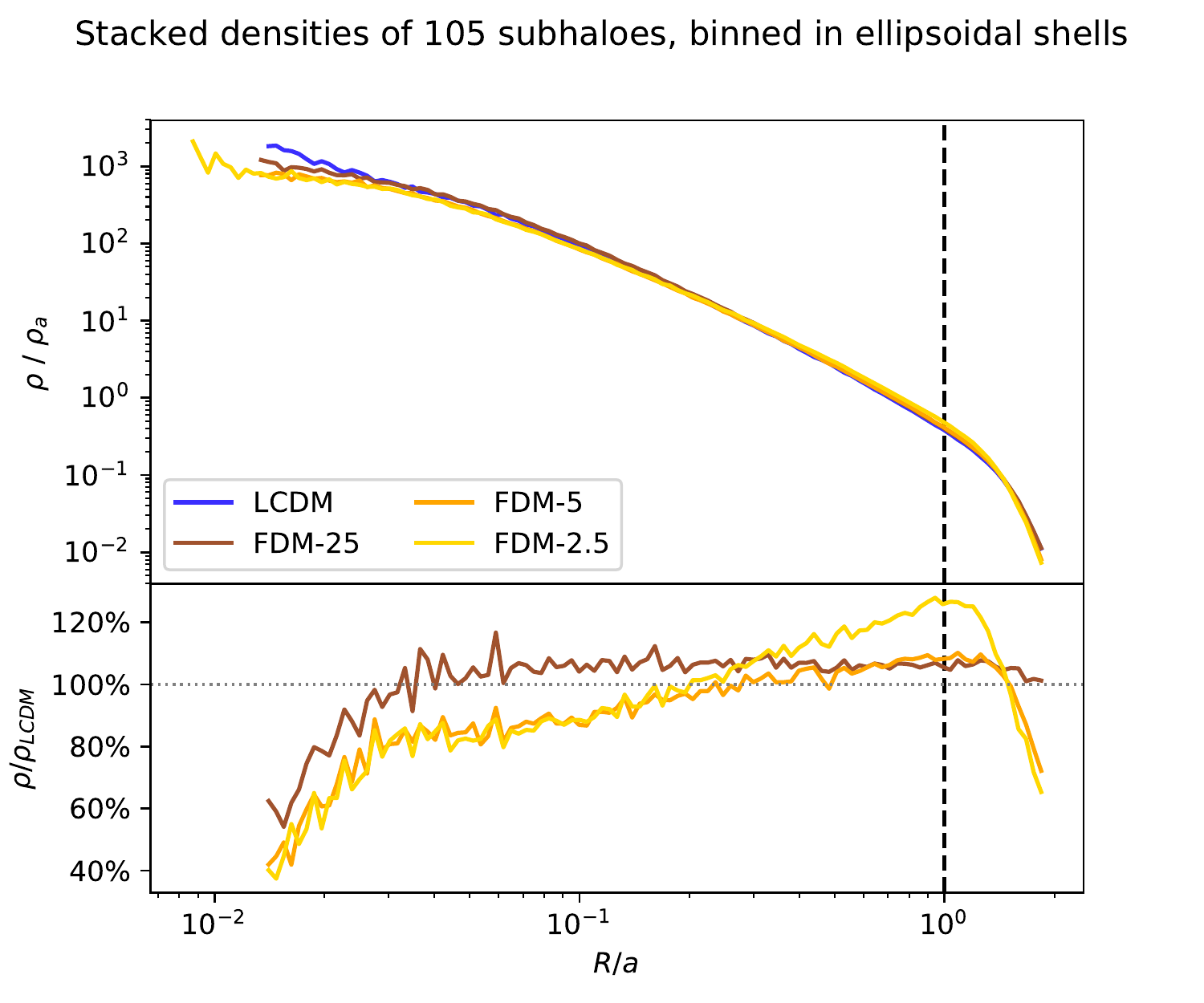} &
\includegraphics[width=0.32\textwidth,trim={0.1cm 0.3cm 1.4cm 1.5cm},clip]{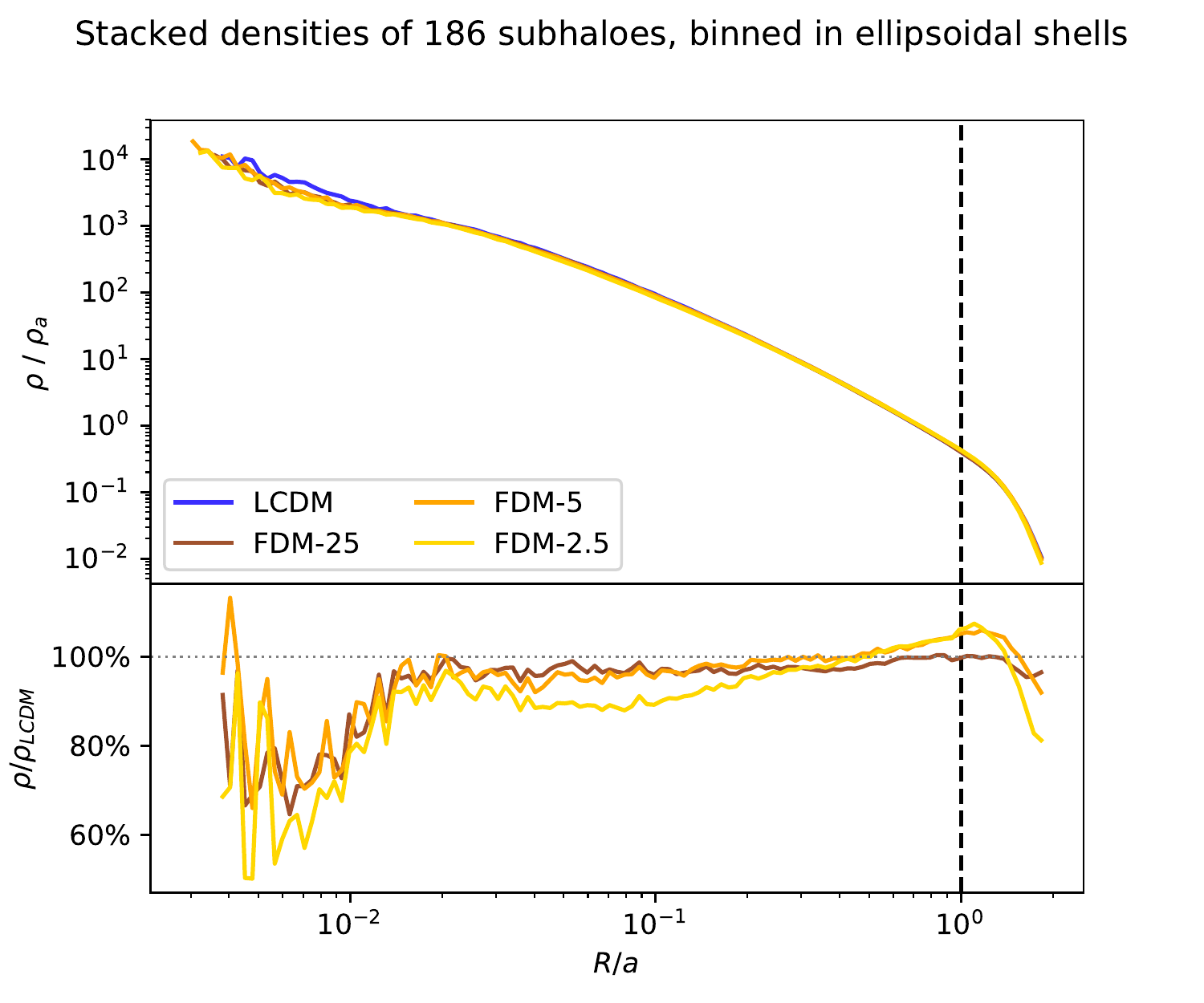} &
\includegraphics[width=0.32\textwidth,trim={0.1cm 0.3cm 1.4cm 1.5cm},clip]{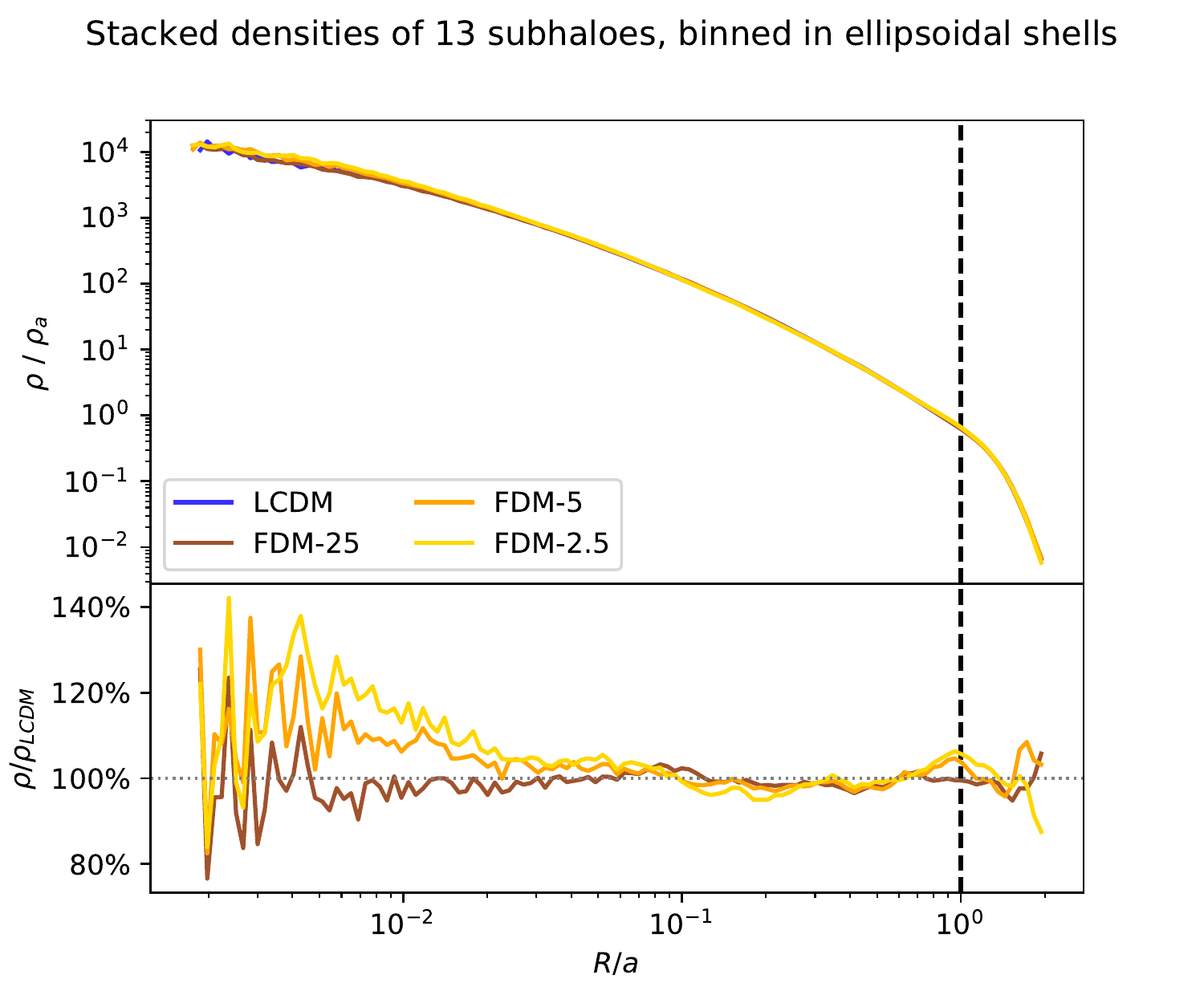} \\
\end{tabular}
\caption{Properties of inter-simulation matching haloes. The total sample is divided column-wise in three mass ranges. The sphericity, the volume occupied and the total mass of the haloes --~contrasted with the corresponding $\Lambda$CDM match~-- are shown in the first row (left panels), together with related distribution functions (right panels). The second and the third row represent the overall density profiles, stacked in fractional spherical shells of $R_{200}$ and ellipsoidal shells of the major axis $a$ --~identified with the vertical dashed lines~--, respectively. Density profiles are divided by the value of the density calculated within $R_{200}$ and $a$ and are shown both in absolute value (top panels) and relatively to $\Lambda$CDM. (bottom panels)}
\label{fig:SAMPLE}
\end{figure*}

Properties of inter-simulation matching haloes are gathered in Fig.~\ref{fig:SAMPLE}, where the total sample is divided column-wise in the three mass ranges. The sphericity, the volume occupied and the total mass of the haloes --~contrasted with the corresponding $\Lambda$CDM match~-- are shown in the first row (left panels), together with related distribution functions (right panels). The second and the third row represent the overall density profiles, stacked in fractional spherical shells of $R_{200}$ and ellipsoidal shells of the major axis $a$ --~identified with the vertical dashed lines~--, respectively, . Density profiles are divided by the value of the density calculated within the $R_{200}$ and $a$ shells and are shown both in absolute value (top panels) and relatively to $\Lambda$CDM (bottom panels).

The sphericity distributions confirm that, in the mass range considered, there is no statistical deviation from $\Lambda$CDM, except for a mild deviation towards less spherical configurations of the less massive haloes, especially in the $m_{22}=2.5$ model. This is consistent with the analysis of the sphericity distributions of the genuine samples (see lower panels in Fig.~\ref{fig:SPHERICITIES}) that reveals that haloes appear to be statistically less spherical with respect to $\Lambda$CDM at $z=0$ when lower FDM masses are considered, down to a maximum of $\sim 10\%$ decrease in sphericity for $m_{22}=2.5$ and halo mass of $\sim 5\times10^{9} M_{\odot}$.

For all the FDM models the volume occupied by the haloes is systematically larger, consistently with a \textit{delayed} dynamical collapse of the haloes. All mass ranges show such property and it is emphasized by lower $m_{22}$ mass --~i.e. stronger QP force~--; however, while bigger haloes occupy almost systematically $20\%$ more volume for $m_{22}=2.5$, smaller haloes can reach even twice the volume occupied by their $\Lambda$CDM counterparts when the same model is considered.

Comparing the mass of the haloes in the various models with the one in $\Lambda$CDM, it is possible to see that small haloes are less massive and big ones, on the contrary, become even more massive, confirming our hypothesis of mass transfer from substructures towards main structures.

The stacked density profiles provide even more insight on the underlying different behaviour between the chosen mass ranges. Starting from the less massive one, the stacked profiles look very differently if plotted using the spherical $R_{200}$-based or the ellipsoidal $a$-based binning. This is due to two concurrent reasons related to the properties of this mass range: first of all, as we said before, the sphericity is $m_{\chi}$ dependent and thus it is not constant with respect to $\Lambda$CDM, so the geometrical difference in the bin shape becomes important when different models are considered; secondly, since the FDM haloes have lower mass but occupy larger volumes, the two lengths are different from each other --~being $R_{200}$ related to density and $a$ purely to geometry~-- so that the actual volume sampled is different. Nevertheless, it is possible to see that in FDM models there is an excess of mass in the outskirts of the halo --~seemingly peaking exactly at distance $a$~-- and less mass in the centre.

The intermediate mass range shows also a suppression in the innermost regions but a less pronounced over-density around $a$ as expected, since the effectiveness of the repulsive force induced by the QP in tilting the density distribution decreases as its typical scale becomes a smaller fraction of the size of the considered objects. In fact, stacked density profiles of the most massive haloes are very similar in the two binning strategies, being $R_{200} \sim a$ and sphericity constant among the various models, and consistent with no major deviation from $\Lambda$CDM, except for a central over-density. It is our opinion, however, that such feature in the very centre of most massive haloes could be a numerical artefact, since its extension is comparable with the spatial resolution used.

The results presented in this Section have been obtained through the detailed analysis of the statistical properties of haloes found at $z=0$ in the FDM simulations. The same analysis, repeated at $z=0$, of the FDMnoQP simulations shows very similar results which are, therefore, not shown in the present work. Such consistency suggests that the properties of haloes at low redshift are --~at the investigated scales~-- not sensible to modifications induced by the dynamical QP repulsive effect, which are expected to appear more prominently at scales of $\sim1 Kpc$ with the formation solitonic cores.

\section{Conclusions}
\label{sec:conclusions}

We have presented the results obtained from two sets of numerical simulations performed with \AG, an extension of the massively parallel N-body code \G for non-linear simulations of Fuzzy Dark Matter (FDM) cosmologies, regarding \LA forest observations and the statistical detailed characterization of the Large Scale Structures.

More specifically, our main aim was to design a set of simulations covering the typical scales and redshifts involved in \LA forest analyses, in order to extract synthetic observations, compare them with  available \LA data, and finally to place a constraint on the mass of the FDM particle. In the literature, \LA forest was already used for this purpose but only in approximated set-ups, in which the quantum dynamical evolution of FDM  was only encoded in the initial conditions transfer function, and neglected during the simulation \citep{Irsic17,Armengaud17,Kobayashi17}, while the \AG code allows us to drop such approximation and take into account the non-linear effects of full FDM dynamics.

The constrain the FDM mass we find is $21.08 \, \times \, 10^{-22}\;\mathrm{eV}$, which is $3\%$ higher with respect to what was found in \citet{Irsic17}. The fact that these two bounds are similar, despite the different dynamical evolution considered in these different works,  implies that the additional suppression deriving from the Quantum Potential dynamical contribution, at the scales and redshifts probed by \LA, is compensated by the gravitational growth of perturbations when these enter the non-linear regime, implying also that --~even if the QP does play a role in the Large Scale Structure evolution~-- the approximation of \citet{Irsic17} \citep[also adopted by][]{Armengaud17,Kobayashi17} is valid and sufficient at these scales.

Secondly, we studied in detail the statistical properties of the Large Scale Structures through the analysis of the aggregated data on haloes regarding their mass, volumes and shapes, as well as their individual inner structure.

\bigskip

The main results regarding the effects of FDM on LSS that we found can be summarized as follows:
\begin{itemize}
\item the FDM particle mass $m_{22}$ defines a typical mass scale $M_t \simeq 1.25 \times 10^{11} / m_{22} \ M_{\odot}$ characterising the halo distribution of different FDM models; all halo properties can be interpreted within the framework of having two families of haloes: the small ones with $M\lesssim M_t$, and the big ones with $M\gg M_t$ (since the very small ones $M\ll M_t$ do not form at all);
\item small haloes, according to the above definition, show outward tilted profiles and a lower total mass, are less spherical and more voluminous, so less dense overall;
\item big haloes instead are almost unaffected in their internal structure --~apart from the expected solitonic inner cores that we cannot resolve with our simulations~--, they occupy a larger volume and they also have higher total mass, mostly accreted outside $R_{200}$, compatible with the collection of the subhaloes mass that were not able to form
\end{itemize}

To conclude, we have performed for the first time a suite of hydrodynamical simulations of a statistically significant  volume of the universe for Fuzzy Dark Matter models featuring a fully consistent implementation of the Quantum Potential effects on the dynamical evolution of the system. These simulations allowed to perform for the first time a fully consistent comparison of mock \LA observations with available data and to update existing constraints on the allowed FDM mass range. As the new constraints are not significantly different from previous ones, this represents the first direct validation of the approximations adopted in previous works. Furthermore, our large halo sample allowed us to perform an extensive characterisation of the properties of dark matter haloes in the context of FDM scenarios, highlighting the typical mass scale below which FDM effects start to appear. Higher resolution simulations will soon allow us to explore even smaller scales where we expect to observe the formation of solitonic cores.

\section*{Acknowledgements}
MN and MB acknowledge support from the Italian Ministry for Education, University and Research (MIUR)
through the SIR individual grant SIMCODE, project number RBSI14P4IH. The simulations described in this work have been performed on the Marconi supercomputer at CINECA thanks to the PRACE allocation 2016153604.
MV and RM are supported by INFN I.S. PD51-INDARK.

%%%%%%%%%%%%%%%%%%%%%%%%%%%%%%%%%%%%%%%%%%%%%%%%%%



\bibliographystyle{mnras}
\bibliography{BIB,baldi_bibliography} % if your bibtex file is called example.bib

%%%%%%%%%%%%%%%%%%%%%%%%%%%%%%%%%%%%%%%%%%%%%%%%%%

%%%%%%%%%%%%%%%%% APPENDICES %%%%%%%%%%%%%%%%%%%%%

\bigskip

\appendix

\section{NOTE ON DE BROGLIE-BOHM INTERPRETATION}
\label{sec:dbb}
In the de Broglie--Bohm (DBB) interpretation of quantum mechanics, the Universe possess at each time a well-defined configuration which evolves under the influence of the wave-function of the system, also know as "pilot-wave". For simplicity and analogy with our problem, let us reduce the Universe configuration to the collective position of N boson particles: in this case the configuration $Q \equiv \left(\vec q_1, \vec q_2, ... ,\vec q_N\right) \in \mathbb{R}^{3N}$ is physically related to the quantum wave-function $\field(Q,t) \in \mathbb{C}^{3N}$.

The wave-function that governs the evolution of $Q$ is the so-called "guiding function"

\begin{equation}
\label{eq:GF}
\frac {d} {dt} \vec q_k = \frac {\hbar} {m_k} \Im{ \frac {\vec \nabla_k \field (Q)} {\field(Q)} }
\end{equation}

while the pilot-wave evolves under the standard Schr\"odinger equation

\begin{equation}
i \hbar \ \de{t} \field(Q) = - \sum_{k=1}^{N} \frac {\hbar^2} {2 m_{\chi}} \nabla_k^2 \field(Q) + V \field(Q)
\end{equation}
.

The DBB is explicitly non-local, given the dependence of velocity of a single particle $k$ on the global wave-function that represents the whole particle ensemble configuration. The Bose-Einstein condensate assumption is then a key ingredient to recover locality that is intimately connected with SPH and its approximation of neighbours-only cut-off. Under condensation, the wave function factorizes

\begin{equation}
\field(Q) \equiv \field(q_1,q_2,...,q_N) = \prod_{k=1}^N \field_k(q_k)
\end{equation}

and, consequently, Eq.~\ref{eq:GF} reduces to Eq.~\ref{eq:madelung} where the dependence on all the other particles different from $k$ cancels in the ratio.

In the DBB framework, the Born Rule $\rho = |\field|^2$ is neither assumed nor even imposed, allowing for \textit{quantum non-equilibrium} states for which this condition is not fulfilled. Yet, it has been shown numerically how a system in which the Born Rule is not initially verified eventually evolve toward quantum equilibrium $\rho \rightarrow |\field|^2$ and, once reached, do not leave \citep{Towler11}. 

Therefore, our answer to the fair question on the consistency of representing a non-local quantum interaction as QP with particle ensembles is that the SPH description of a bosonic FDM specie in the Bose-Einstein condensation regime is theoretically robust and coherent with wave-based portrayals. In \citet{Nori18} we showed how SPH is indeed able to recover some of the FDM results obtained with full-wave solvers.

It is nevertheless important to stress that the equivalence between the Eulerian and Lagrangian pictures do not ensure overlapping results in terms of numerical simulation, since the intrinsic temporal and spatial resolution is finite and affects the two differently. For these reasons, however, it is our belief that incompatibilities between the two approaches are to ascribe only to resolution limits.

In this sense it is very interesting the convergence to the classical results in the limit $\hbar / m \rightarrow 0$ shown in \citet{Mocz18}, where the potential and the force in the Schr\"odinger--Poisson description obtained by simulations in several tests approach the classical Vlasov--Poisson ones while the density field is however unable to do the same due to uncontrollable interference patterns. 

Studying the accuracy and the behaviour in limit cases of numerical realizations of quantum systems not only is necessary to estimate the deviation between simulations and observations but can be useful to improve our understanding of statistical representations of quantum nature objects.

%%%%%%%%%%%%%%%%%%%%%%%%%%%%%%%%%%%%%%%%%%%%%%%%%%

% Don't change these lines
\bsp	% typesetting comment
\label{lastpage}
\end{document}